# Urban wireless traffic evolution: the role of new devices and the effect of policy

Jaume Benseny[1], Jarno Lähteenmäki[1], Juuso Töyli[1,2], Heikki Hämmäinen[1]

[1]Communication and Networking Department, Aalto University,
Konemiehentie 2, P.O.Box 15400, Espoo, Finland

[2]Department of Economics, University of Turku,
Rehtorinpellonkatu 3, 20500, Turku, Finland

jaume.benseny@alumni.aalto.fi, juuso.toyli@utu.fi
{jarno.lahteenmaki, heikki.hammainen}@aalto.fi

## Abstract

The emergence of new wireless technologies, such as the Internet of Things, allows digitalizing new and diverse urban activities. Thus, wireless traffic grows in volume and complexity, making prediction, investment planning, and regulation increasingly difficult. This article characterizes urban wireless traffic evolution, supporting operators to drive mobile network evolution and policymakers to increase national and local competitiveness. We propose a holistic method that widens previous research scope, including new devices and the effect of policy from multiple government levels. We provide an analytical formulation that combines existing complementary methods on traffic evolution research and diverse data sources. Results for a centric area of Helsinki during 2020-2030 indicate that daily volumes increase, albeit a surprisingly large part of the traffic continues to be generated by smartphones. Machine traffic gains importance, driven by surveillance video cameras and connected cars. While camera traffic is sensitive to law enforcement policies and data regulation, car traffic is less affected by transport electrification policy. High-priority traffic remains small, even under encouraging autonomous vehicle policies. Based on peak hour results, we suggest that 5G small cells might be needed around 2025, albeit the utilization of novel radio technology and additional mid-band spectrum could delay this need until 2029. We argue that mobile network operators inevitably need to cooperate in constructing a single, shared small cell network to mitigate the high deployment costs of massively deploying small cells. We also provide guidance to local and national policymakers for IoT-enabled competitive gains via the mitigation of five bottlenecks. For example, local monopolies for mmWave connectivity should be facilitated on space-limited urban furniture or risk an eventual capacity crunch, slowing down digitalization.

## Keywords

**Abbreviations**

IoT : Internet of Things, mmWave : millimeter wave, M2M : machine-to-machine, POS: Point of sale, CAGR: Compound annual growth rate.

1. **Introduction**

Wireless technologies accounted for 10% of the global income per capita growth of the last two decades, mainly included in human computing devices, such as smartphones. More recent wireless technology, such as the Internet of Things (IoT), can diffuse into all kinds of products like cars, watches, etc. In the next 10 years, IoT is expected to enable 2.1% of global income increase, playing an important role in productivity growth (GSMA, 2020). Whether countries can leverage IoT technologies for higher productivity and competitiveness depends on the prompt availability of IoT connectivity, i.e., their technological readiness, and the ability of application developers and end-users to maximize value creation, i.e., their technological innovation capabilities.

IoT technologies can bring a plethora of new devices on-line, introducing new traffic patterns that differ from broadband. IoT devices currently present modest volumes, limited mobility, and highly periodic activity, predominantly via uplink rather than downlink. However, IoT applications are rapidly developing in advanced markets, where industries observe growing per device volumes and expanding mobility (Finley et al., 2020; L. Liu et al., 2022; Rathore et al., 2016; M. Shafiq et al., 2020; Tahaei et al., 2020). Therefore, urban wireless traffic prediction becomes increasingly difficult, raising uncertainty for operators and telecom policymakers.

In contrast to wireless broadband services, IoT adoption is not only driven by consumers and enterprises but also by governments, which may directly purchase or encourage its usage (Hatuka & Zur, 2020; Ylipulli & Luusua, 2020). Government policies across administrative levels increasingly rely on wireless technologies to support public services. In advanced markets, detailed public transport information is widely available thanks to street traffic and vehicle monitoring. Smart parking applications inform in real-time about availability of parking space. On energy, street lightning is being automated, and electricity metering is periodically shared. On law enforcement, street video surveillance networks have been deployed in some cities to monitor crime and terrorist threads (Caragliu & Del Bo, 2019; Lin et al., 2017; Neirotti et al., 2014). Hence, IoT can change the structure of urban wireless traffic with implications for operators, including revenues from new devices and investments to enable new connectivity features.

The fifth generation of mobile technology (5G) has the potential to make IoT widely available, thus providing connectivity to a new range of markets compared to previous mobile technologies. Therefore, it has been suggested that 5G policy adopts an adaptive form with focus on the creation of guidelines for market players, given the higher than usual market uncertainty (Blind & Niebel, 2022; Lehr et al., 2021). Boundary conditions for market and regulatory designs may encourage new players (i.e., other than mobile network operators) to explore industrial, government, and community applications (Bauer & Bohlin, 2022). For novel applications to emerge, varying throughput and latency requirements should be satisfied by combining new virtual components with appropriate frequencies (Knieps & Bauer, 2022). In Europe, 700 MHz, 3.5 GHz, and 26 GHz were defined as 5G pioneering bands. While the first two allow incremental evolution of mobile networks, the latter may require architectural changes, potentially leading to alternative connectivity business models (Benseny et al., 2020; Lehr et al., 2021). This work aims to understand when traffic demand may need these frequencies and discuss regulatory options for its adoption via small cell architectures.

The evolution of wireless traffic has been studied, including the renewal of device populations (Kivi et al., 2012; Riikonen et al., 2013), the diffusion of wireless technologies (Jha & Saha, 2020; Lim et al., 2012; Nam et al., 2008), the forecasting of country-level volumes (Cisco, 2020; Ericsson, 2019b; Lee et al., 2016), and the techno-economic assessment of new wireless services (Ajibulu et al., 2017; Bao et al., 2017; Schneir et al., 2019). However, a broader research approach to wireless traffic evolution is now required, given the advent of new devices and the effect of policy from multiple government levels.

This article aims to reveal changes in urban wireless traffic, supporting operators to drive mobile network evolution and policymakers to increase national and local competitiveness. Changes may not be sufficiently captured by existing methods, given the advent of IoT devices with different traffic behavior and policy effects from more government levels. The effect of policy in this article is two-fold. Initially, policy effects are considered as inputs to test the sensitivity of traffic to policies from multiple government levels across different domains. Later, telecom and local policy are discussed for mobile network evolution, leading to rich IoT adoption and competitiveness gains.

To achieve the article's aim, we first identify the fundamental processes underlying evolution and we propose a holistic method for its investigation. We provide an analytical formulation to estimate daily volumes of user traffic and indicators of control traffic. User traffic is the fraction of traffic conveying user content, and control traffic is the fraction that includes network signaling. Second, we define device categories, addressing the needs of personal connectivity and wellbeing, transport and mobility, urban sensing and safety, and retail logistics. Third, we identify policies from multiple government levels that can affect the adoption and usage of these device categories. Fourth, we combine complementary methods previously used in the study of traffic evolution and diverse data sources. Finally, we estimate traffic evolution for a Helsinki city center's postal code until 2030, considering slow and rapid demand growth scenarios, covering for uncertainty. Helsinki is a relevant study case, given the leadership of Finland in wireless service adoption. This Nordic country is among the top 5 machine-to-machine (M2M) connection penetration and the first in mobile data usage per subscription. Moreover, the urban densities of Helsinki match those of typical medium-sized European cities.

The rest of this article is structured as follows. Section 2 reviews previous work on wireless traffic evolution. Section 3 presents a method for investigating urban wireless traffic evolution. Section 4 presents the employed data, including the Helsinki area under study. Section 5 uncovers results for penetration, user volumes, control needs, and urban wireless traffic evolution. Section 6 discusses the effect of policy on traffic evolution, including demand side effects and implications for telecom and local policy. In Section 7, conclusions are drawn.

## 2. Previous work

Previous work has studied multiple facets of the traffic evolution, uncovering interdependencies between the development of wireless device populations and the usage of its associated applications.

Wireless traffic depends on the features of the device population since they define the device usage limits. Studies on smartphone usage agree that new devices, which are usually equipped with the latest features, generate higher-than-average traffic (Jin et al., 2012; Malandrino et al., 2017). From a method viewpoint, the evolution of device populations has been studied through replacement purchase models (Olson & Choi, 1985) and technological substitution models (Fisher & Pry, 1971). For mobile handsets, it has been characterized by combining demand-driven product replacement and supply-driven product feature dissemination, including wireless technology, display, camera, and positioning technology, among others (Kivi et al., 2012; Riikonen et al., 2013).

The diffusion of wireless technology into device populations enables more advanced applications. The diffusion of new wireless technologies in parallel with the adoption of new applications was retrospectively studied for the case of Canada, explaining traffic volume growth (Halepovic et al., 2009). Wireless technology diffusion has been commonly studied via contagious models, characterized by an S-shape curve (Meade & Islam, 2006). For example, the adoption of 2G mobile subscriptions in China was modeled via a Bass model and used to forecast 3G penetration (Lim et al., 2012). A similar approach is employed to estimate future demand for the WiBro service in Korea (Nam et al., 2008). Recently, the timeliness of 5G service adoption has been researched via an ordered logit model, identifying cost and lack of need as significant purchase delayers (Maeng et al., 2020).

The development of application usage increases the traffic volume of devices. Studies based on smartphone measurements indicate a growing trend in daily smartphone usage, which has reached 3 hours in some European countries. It is generally accepted that video-based applications are the most traffic-intensive and that the traffic distribution with respect to applications is highly skewed (X. Liu et al., 2018; Walelgne et al., 2021). In this context, traffic has been forecasted at the country-level considering multiple devices and applications. A 10-year forecast for Korean mobile broadband traffic was conducted via a three-round Delphi process, considering penetration, daily time usage, and application usage for multiple devices (Lee et al., 2016). Wireless equipment vendors, such as Ericsson and Cisco, periodically release country-level traffic forecasts with a time horizon of 5 to 6 years, including broadband and IoT devices. However, their methods are not known.

Traffic characteristics vary depending on the urban area. Investigations on base station data indicate that urban traffic can be satisfactory modeled, considering the socio-economic activity of the coverage area, including the following area types: residential, entertainment, transport, and office (Cici et al., 2015; Ding et al., 2018; Xu et al., 2017). Traffic forecasts exist for delimited urban areas supporting the techno-economic evaluation of new 5G services (Ajibulu et al., 2017; Bao et al., 2017; Schneir et al., 2019). These studies estimate user and control traffic, based on urban features, i.e., user densities from central London, and typically including IoT devices. For example, results for a multi-device case indicate that daily volumes in 2030 reach 1.6 PB/day or 42 TB/km2 dominated by broadband traffic from consumer devices, followed by vehicle infotainment. However, these work limit traffic evolution to the boundaries established by the studied technology. As a result, it appears that long-term evolution is heavily influenced by static protocol and service specifications rather than dynamic actor needs or government policies.

## 3. Method

### 3.1. Evolution of urban wireless traffic

The evolution of wireless traffic is a complex process driven by digitalization, which expands the wireless device population, widens the range of communication needs, and extends the usage of applications. This process is enabled by the substitution of competing wireless technologies and applications, which overtime allow wireless systems to satisfy a broader range of needs. When a new technology becomes available, it can diffuse into the device population, relaxing the hardware limits for device usage. For example, new mobile technologies enable smartphones to reach higher download speeds. Another example is the diffusion of high-definition screens, which allow higher bitrates for video streaming. As urban activities get digitalized, new application features can also diffuse into the population of installed applications, enabling additional and enhanced functions. For example, connected car applications can be extended to inform about road or traffic conditions. Finally, the usage of individual applications also develops, as users extend usage to other domains and expand mobility to other locations. As a result, devices generate more user and control traffic, leading to increasing aggregated volumes. The speed of these underlying processes is respectively established by the costs of hardware development, software development, and user time. Figure 1 describes the proposed characterization of urban wireless traffic evolution.

Page **5** of **44**

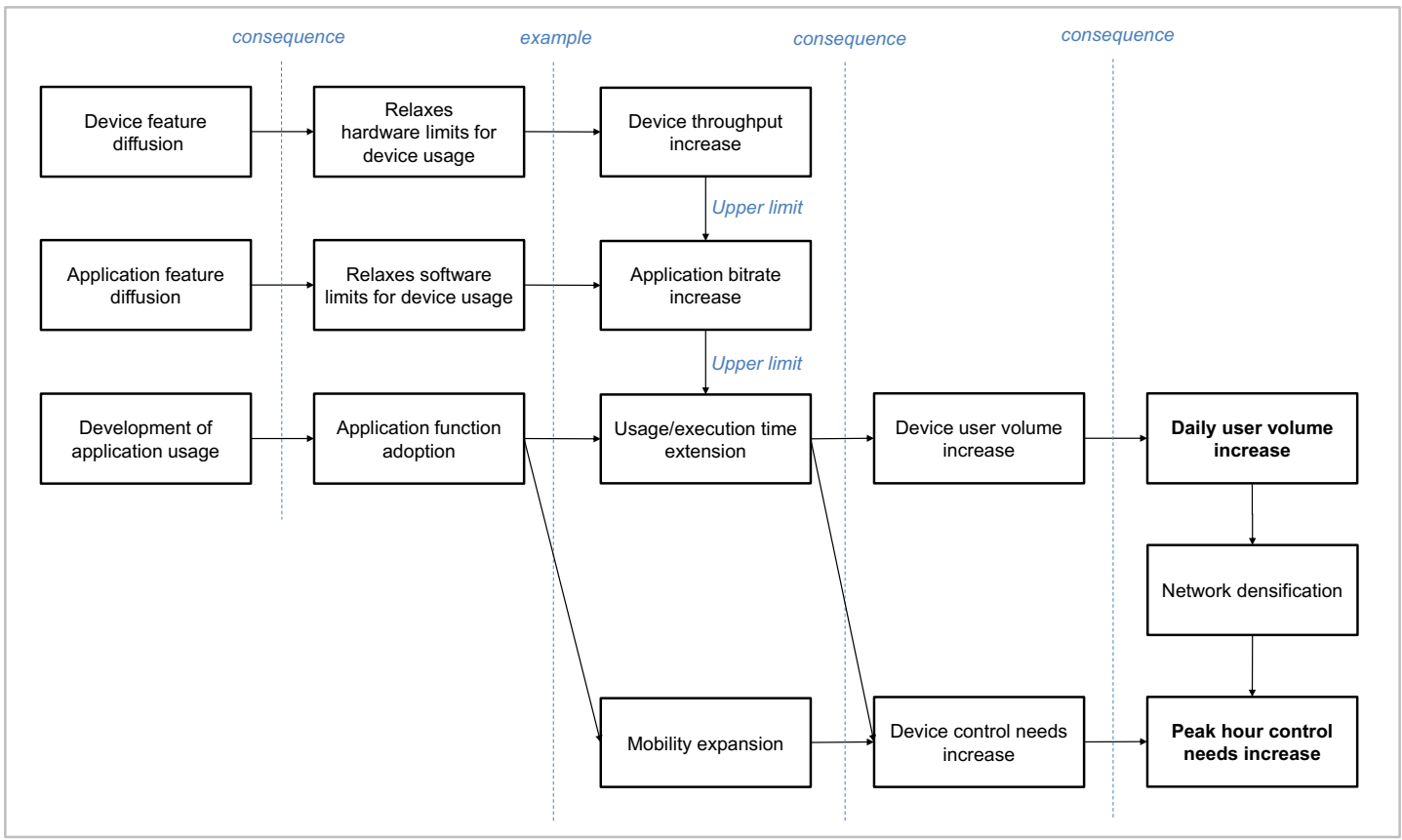

*Figure 1 Processes underlying the traffic evolution*

At a higher level of abstraction, the digitalization process is conducted by urban actors, who purchase and use devices and applications to address individual and societal needs. These actors' behavior can be influenced by policies from multiple government levels, accelerating or slowing down the processes underlying the traffic evolution.

### 3.2. Overview

We propose a holistic method to study the evolution of the user traffic and the control traffic between devices and network access points in urban areas. Specifically, the method estimates the evolution of daily volumes for user traffic, including 0-23h distribution, and key indicators for control traffic at peak hour. The method estimates average daily volumes, and it does not aim to estimate instantaneous peaks.

First, we define 11 device categories, including smartphones, mobile modems, wearables, connected bikes, connected cars, traffic signs and lights, autonomous buses, street surveillance video cameras, urban sensors, electricity and water meters, parcel delivery drones, and point of sale (POS) devices. We characterize the diffusion of wireless technology into a device category via the proxy variable *device penetration* (See step 1 in Figure 2). For example, we characterize the diffusion of mobile technology into cars through connected car penetration. Second, we estimate *device density* by associating *device penetration* to key urban densities, which fluctuate during 0-23h (see step 2). Third, we characterize the diffusion of application features together with the development of application usage through two proxy variables, namely *application user volumes* and *application control needs* (steps 3a and 3b). The former informs about the evolution of device user volumes and the latter on the evolution of device control needs. This distinction is needed because network capacity is typically dimensioned accounting for both user and control traffic. Importantly, IoT may significantly increase the fraction of traffic dedicated to control functions compared to traditional broadband, given the different behavior of IoT devices. Hence, we define control needs as the intermittent attachment of devices with low traffic activity and the connection handover of moving devices between base stations. Fourth, we estimate the daily user volume as the sum of *device user contributions* that result from combining *device density* and *device user volume* (step 4). *Device user volume* is the sum of application contributions for this device, according to application-to-device correspondences, defined in a later section. Fifth, we estimate peak hour control needs as the sum of *device control contributions* that result from *device density* and *device control needs* (step 5). Finally, to account for uncertainty, including the effect of policy on the purchase and usage of devices and applications, we generate slow and rapid demand growth scenarios. The slow scenario includes low estimates for *device penetration*, *application user volume*, and *application control needs* during the study period. The rapid scenario combines the corresponding high values.

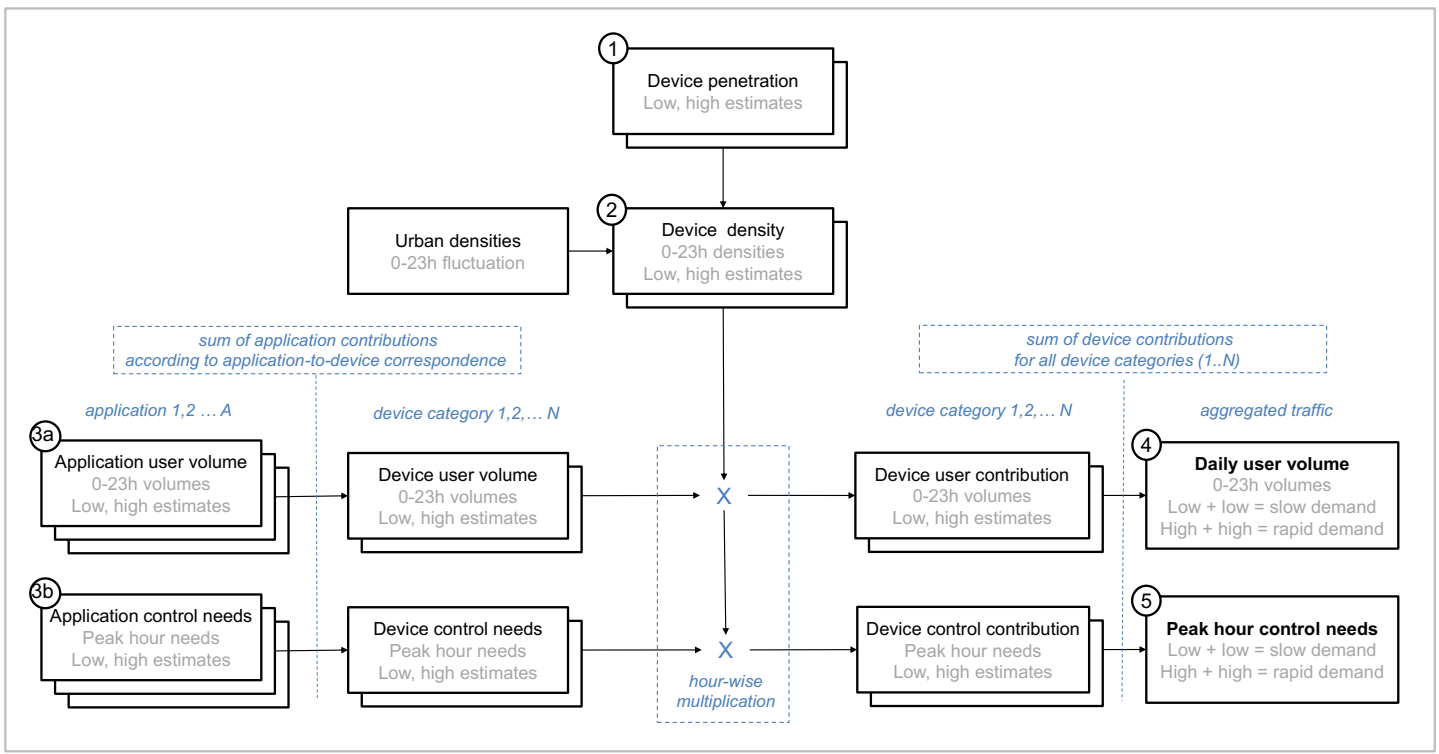

*Figure 2 Method overview*

### 3.3. Analytical formulation

Device density

For a given year, we define the 0-23h distribution of the device density for the device *i* via the array $\vec{d}_{i,year}$ with dimensions 1x24, where array positions correspond to hours of the day. We calculate the values for this array as follows:

$$\vec{d}_{i,year} = p_{i,year} \cdot \vec{u_i} \qquad \text{(Eq. 1)}$$

where $p_{i,year}$ is the annual penetration of the device *i* with respect to its adopting body. For smartphones, the adopting body is the population. Moreover, $\vec{u_i}$ is the urban density associated with the device during 0-23h. For example, to estimate the smartphone density, smartphones can be associated with the active population density.

The annual evolution of the device density is considered via the development of device penetrations. Penetration levels follow an S-shaped curve until the maximum carrying capacity of the adopting body is reached. In this work, we use a combination of pre-existing methods and models depending on the device. We use the Bass model when historical data is available; a replacement purchase model for devices with data on population, sales, and lifetime; a coverage-based models for devices with geographically constrained applications; and exceptionally device-specific models.

The future purchase of devices can hardly be predicted without error. Hence, we model the device penetration (i.e., $p_{d,year}$) to account for uncertainty on use case needs, including the effect of government policies. For example, connected car sales can increase if transport electrification policies accelerate the car stock's renewal via subsidies. Hence, we establish that values of $p_{d,year}$ are limited between a low and a high boundary, as follows:

$$p_{i,year,low} < p_{i,year} = f(use\ case\ needs\ |\ policy) < p_{i,year,high} \qquad \text{(Eq. 2)}$$

Daily user volume

For a given year, we define the 0-23h distribution of the daily user volume $\vec{w}_{year}$ through the following element-wise array multiplication:

$$\vec{w}_{year} = \sum_{i=1}^{N} (\vec{d}_{i,year} * \vec{t}_{i,year}) \qquad \text{(Eq. 3)}$$

where *N* is the number of devices, and $\vec{t}_{i,year}$ is the user volume generated by the device *i* during 0-23h. In addition, we define $\vec{t}_{i,year}$ as:

$$\vec{t}_{i,year} = \sum_{a=1}^{A}\left(\vec{c}_{a,i,year,category}\right) \tag{Eq. 4}$$

where $A$ is the number of applications that generate traffic for the device $i$ and $\vec{c}_{a,i,year,category}$ is the user volume generated by the application $a$, which belongs to a specific traffic category.

We study the year-on-year evolution of the device user traffic by estimating the volume growth of individual application (i.e., $\vec{c}_{a,i,year,category}$) by employing the following growth functions: exponential, ceiling-limited linear, and a constant value. Also, we combine publicly available traffic volumes as well as trends and forecasts from well-known third-parties.

Future application volumes cannot be predicted easily. Hence, we model $\vec{c}_{a,d,year}$ to account for use case uncertainty, including the effect of policies, as follows:

$$\vec{c}_{a,i,year,low} < \vec{c}_{a,i,year} = f(use\ case\ needs\ |\ policy) < \vec{c}_{a,i,year,high} \tag{Eq. 5}$$

Summarizing, the equation for the rapid demand scenario is shown as follows:

$$\vec{w}_{year,rapid} = \sum_{i=1}^{N}\left(\left[p_{i,year,high} \cdot \vec{u_d}\right] * \sum_{a=1}^{A}\left(\vec{c}_{a,i,year,high}\right)\right) \tag{Eq. 6}$$

Peak hour control needs

We define the rate of attachment requests and the rate of handover requests as indicators for the aggregated control needs at peak hour. For a given year, we define the attachment rate $r_{year}$ as:

$$r_{year} = \frac{1}{t_{r,min}} \sum_{i=1}^{M}\left(d_{i,year,peakhour} \cdot \alpha_{i,year}\right)\ ;\ \alpha_{i,year} = \frac{1}{t_{r,i,year}} \tag{Eq. 7}$$

where $M$ is the number of low-activity devices, $t_{r,min}$ is the minimum inter-request time among these devices, and $\alpha_{j,year}$ is the probability of attachment for the device $i$ during the length of $t_{r,min}$. This probability is calculated based on the inter-request time of devices in hours. In addition, we define the handover rate $h_{year}$ as:

$$h_{year} = \frac{1}{t_{h,min}} \sum_{i=1}^{Q}\left(d_{i,year,peakhour} \cdot \beta_{i,year}\right)\ ;\ \beta_{i,year} = \frac{t_{h,min} \cdot s_i}{l_{year}} \tag{Eq. 8}$$

where $Q$ is the number of high-activity devices and speeds greater than zero, $t_{h,min}$ is the travel time of the fastest device when traversing the site coverage diameter, and $\beta_{i,year}$ is the probability of handover for the device *i* during $t_{h,min}$. This probability is calculated based on the speed of devices $s_i$ and the inter-site distance $l_{year}$.

We study the year-on-year evolution of the attachment rate by considering the development of the inter-request time (i.e., $t_{r,i,year}$) for low traffic profile devices. For handover rates, we consider the development of the average inter-site distance (i.e., $l_{year}$) for the wireless network. We estimate the annual development of device inter-attachment rates and network inter-site distances via bottom-limited decreasing linear functions or a constant value. To account for uncertainty on year-on-year evolution of these two rates, we estimate boundary values for inter-request times and inter-site distances.

### 3.4. Assumptions

We suggest that citizens, enterprises, and municipalities purchase and use devices as well as applications to satisfy needs in personal connectivity and wellbeing, transport and mobility, urban sensing and safety, and retail logistics. In Table 1, we identify key devices and associated applications, given trends in digitalization policies (Caragliu & Del Bo, 2019; Grant-Muller & Usher, 2014; Kyriazis et al., 2013; Neirotti et al., 2014), observations from operator network data (Finley et al., 2020; Romirer-Maierhofer et al., 2015; M. Z. Shafiq et al., 2013), forecasts from network device manufacturers (Cisco, 2020; Ericsson, 2020), and forward-looking views from academic researchers (Batty et al., 2012; Khan et al., 2015; Rathore et al., 2016). Also, we classify applications into traffic categories described in detail in Annex 0.

*Table 1 Societal needs, urban actors, applications, and device categories*

| Societal need | Device category | Application | Traffic category | Purchaser / user |
|---|---|---|---|---|
| Personal connectivity and wellbeing | Smartphones and mobile modems | Multiple via app. market | (1) Human traffic | Citizen |
| Personal connectivity and wellbeing | Wearables | Multiple via app. market | (2) Machine low-activity traffic | Citizen |
| Transport and mobility | Connected bikes | Tracking | 2 | Citizen |
| Transport and mobility | Connected cars | Remote monitoring | (3) Machine high-activity traffic | Citizen |
| Transport and mobility | Connected cars | Multiple via app. market (Infotainment) | 1 | Citizen |
| Transport and mobility | Traffic signs and lights | Sensing and acting | 2 | Municipality |
| Transport and mobility | Autonomous buses | Remote driving | (4) High-priority traffic | Municipality |
| Transport and mobility | Autonomous buses | Remote monitoring | 3 | Municipality |
| Urban sensing and safety | Street surveillance video cameras | Video streaming and processing | 3 | Municipality |
| Urban sensing and safety | Urban sensors[1] | Sensing and acting | 2 | Municipality |
| Urban sensing and safety | Electricity and water meters | Sensing and acting | 2 | Enterprise |
| Retail logistics | Parcel delivery drones[2] | Remote monitoring | 3 | Enterprise |
| Retail logistics | POS devices | Payment, logistics | 2 | Enterprise |

[1] For urban sensors, we mean low battery, low activity sensors that periodically transmit measurements on air quality sensors, bridge vibration, trash bin levels, etc.
[2] The current use of drones for inspection and surveillance (e.g., street traffic surveillance, building and infrastructure inspection, police support) may not significantly affect urban wireless traffic. The effect can become significant when its commercial use sees massive adoption, for example, for parcel delivery.

Table 2 presents assumptions and supporting methods for the estimation of traffic evolution. To estimate penetration, it presents adopting bodies and estimation methods. For device density, it associates devices to urban densities and their 0-23h dynamics. For device user traffic, it presents volume growth functions. For connected cars and autonomous buses, we use two functions, one for each application. Note that we do not estimate user volumes for devices that generate traffic belonging to the machine low-activity traffic category, given their very low daily volumes. We refer to tracking applications for connected bikes; payment and logistics applications for POS devices; sensing and acting applications for connected traffic lights and signs; urban sensors, electricity and water meters, and wearables. Nevertheless, we do estimate the device density of these devices (hereafter low-activity devices, see Annex 1) since they are the main contributors to control traffic concerning network attachment. Table 2 also indicates what devices are included to estimate the control traffic indicators.

*Table 2 Assumptions and supporting methods*

| Devices | Adopting body | Penetration method | Urban densities (0-23h dynamics) | User traffic method | Control traffic indicator[2] |
|---|---|---|---|---|---|
| Smartphones | Population | Bass model | Active population density[1] (dynamic) | Exponential growth | Handover rate |
| Mobile modems | Population | Bass model | Non-working population density (dynamic) | Exponential growth | - |
| Wearables | Population | Replacement model | Active population density (dynamic) | - | Attachment rate |
| Connected bikes | Bikes in use | Replacement model | Moving bike density (dynamic) | - | Attachment rate |
| Connected cars | Cars in use | Replacement model | Moving car density (dynamic) | Infotainment: Exponential growth; Remote monitoring: ceiling-limited linear growth | Handover rate |
| Autonomous buses | Buses in use | Replacement model | Moving bus density (dynamic) | Remote driving: Constant value; Remote monitoring: Ceiling-limited linear growth | Handover rate |
| Traffic signs and lights | Postal code area | Coverage-based method | Postal code area (static) | - | Attachment rate |
| Urban sensors | Postal code area | Coverage-based method | Postal code area (static) | - | Attachment rate |
| Street surveillance video cameras | Postal code area | Coverage-based method | Postal code area (static) | Ceiling-limited linear growth | - |
| Electricity/water meters | Buildings | Device-specific method | Building density (static) | - | Attachment rate |
| Parcel delivery drones | Population | Coverage-based method | Active population density (dynamic) | Ceiling-limited linear growth | Handover rate |
| POS devices | Retailers | Device-specific method | Retailer density (static) | - | Attachment rate |

[1] The active population density includes population changes through the hours of the day. For smartphone handovers, we employ a subset of the active population density, which we name the moving population density. Difference is described in the next section.
[2] As mentioned in the formulation, we estimate the annual development of device inter-attachment rates and network inter-site distances via bottom-limited decreasing linear functions or a constant value

The purchase and use of devices and applications are affected by policies, including transport and mobility, law enforcement, digital economy, buildings, aviation, and telecommunications. In Table 3, we indicate what policies affect what devices and whether the effect is on device penetration, device user volume, or both.

*Table 3 The effect of policies*

| Policy domain | Policies / Regulations | Limiting conditions | Effect on |
|---|---|---|---|
| Transport and mobility | Sustainable mobility policy | Bike lane construction, car parking reduction | Device penetration<br>• Connected bikes |
| | Transport electrification policy | Road pricing for polluting vehicles, electric vehicle subsidies | Device penetration<br>• Connected cars<br>• Autonomous buses |
| | Autonomous vehicle regulation | Driver-less vehicle requirements | Device penetration<br>• Autonomous buses<br>• Connected traffic lights and signs |
| Law enforcement | Street video surveillance policy | Street area coverage ambition | Device penetration<br>• Street surveillance video cameras |
| Digital economy | Data protection regulation | Limited processing and storage of personal information | Device user volume<br>• Street surveillance video cameras e.g. video-based applications |
| Building | Electricity and water meters | Meter function requirements | Device penetration<br>• Electricity and water meters |
| Urban monitoring | Air quality policy<br>Waste mgmt. policy<br>Parking policy | Street area coverage ambitions | Device penetration<br>• Urban sensors |
| Aviation | Unmanned aviation regulation | Drone operator requirements, no-fly zones, landing station requirements | Device penetration & user volume<br>• Autonomous drones |
| Telecommunications | Spectrum policy | Support for low-latency and reliable connectivity (device penetration)<br>Additional capacity (device user volume) | All devices |

### 3.5. Scenarios and COVID19 uncertainty

Government-imposed restrictions during the COVID19 pandemic[1] introduced an alteration to demand evolution, changing trends in device sales, daily user commutes, application usage, etc. Since restrictions were temporarily imposed, trend changes are not likely to persist in the long run, except for remote working. To include short-lived changes into our forecasting method could lead to errors when targeting long-time horizons such as 2030. To prevent this, our forecasting only includes pre-COVID19 data.

---

[1] In Finland, the state of emergency was declared in March 2020, including the shutdown of schools and government-run public facilities, the limitations of public meetings to 10 people, and more. Although the nationwide recommendation to work from home was lifted in February 2022, commuting patterns have not returned to pre-pandemic levels, and significant remote working is likely to persist.

We address uncertainty on urban wireless traffic evolution by generating slow and rapid demand scenarios. While the rapid scenario includes high estimates for device density, user volumes, and control needs, the slow scenario consists of the corresponding low values, as shown in Figure 2. To generate low and high estimates, we identify the primary source of uncertainty on the purchase and usage of devices, which can be either the effect of policy or related to the use case (e.g., technical, business uncertainties). The exact rationale behind the estimates is described for each device in the upcoming sections. In three exceptional cases, we generate a medium estimate included in both slow and rapid scenarios. The first two cases include the penetration of smartphones and mobile modems, given their current high penetration levels over potential users. The third case is about electricity/water meters since evidence exists on smart metering policy's effectiveness, e.g., Finland (Tractebel, 2019). Scenarios resulting from low-high and high-low combinations are calculated but not presented due to space limitations.

In summary, the followed approach deals with forecasting uncertainty by providing results within a range of acceptable confidence. Upper and lower bounds for this range correspond with volume estimates for rapid and slow demand scenarios, respectively. These scenarios account for variation in the adoption and usage of technology depending on national/local policy and device-specific factors. The approach does not account for changes in the technology itself. Nevertheless, changes in technology are less likely than in its usage, especially in highly standardized and patent-protected technologies such as 5G. Other limiting factors include the employed forecasting methods, e.g., Bass model, the early life cycle stage of the investigated technology, and the length of the forecasting period.

4. **Data**

We employ pre-COVID19 data to conduct the forecast until 2030, as justified in the previous section.

   4.1. **The case of Helsinki**

Helsinki is an interesting study case, given the leadership of Finland in wireless technology adoption. Therefore, changes in the structure of the urban wireless traffic may occur earlier than in other cities. Importantly, results should have generalization power since Helsinki's urban densities at least match those of medium-sized European cities, in contrast to singular mega-cities or city-states. In 2017, Finnish enterprises presented the highest ICT usage in OCDE countries, and IoT devices are already being adopted across many industries. Finland is among the top 5 countries globally on M2M penetration and the first in mobile data usage per subscription (Finley et al., 2020; OECD, 2017a, 2017b).

Since Finland belongs to the European Union (EU), traffic demand can be affected by policies at the European-level, e.g., General Data Protection Regulation (GDPR), at the state level, e.g., building regulation on smart metering, and at the city-level, e.g., sustainable mobility policy. In this context, Finland has pioneered adopting digitalization policies. For example, the EU Directive in Energy Performance of Buildings was promptly embraced, and the deployment of first-generation smart electricity meters was already completed by 2013 (Tractebel, 2019). Another example is the development of unmanned aviation regulation for commercial drone operations, including the issue of operator licenses. In 2015, the first drone-based package delivery service was trialed for 3 days by the Finnish postal service. In 2020, the Google subsidiary, named Wing, started a food delivery service for citizens in a relatively populated neighborhood (Yleisradio, 2019).

### 4.2. Urban densities

We estimate the wireless traffic evolution for the postal code with the name "Helsinki Keskusta - Etu-Töölö", which concentrates the highest number of workplaces and includes the most commercial streets in Helsinki. Figure 3 (left) shows the postal code location, indicating the workplace density of surrounding postal codes. This figure (right) also shows the street network and land uses of the postal code in detail. We select this area because outdoor traffic volumes per square kilometer should be highest since it covers most of the daily economic activity.

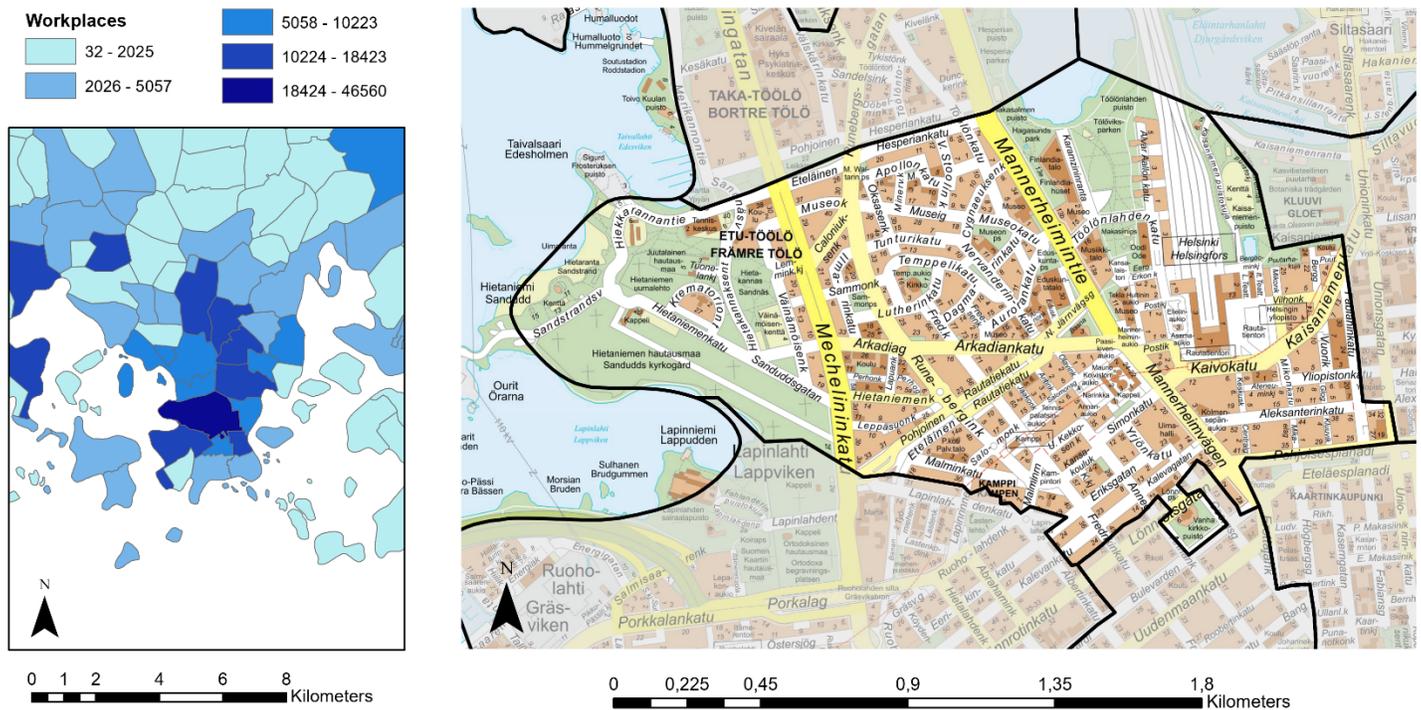

*Figure 3 Postal code understudy*

Next, we describe how urban densities were generated based on multiple Helsinki data sources. Urban densities were presented in Table 2, respectively associated with devices. Although remote working reduced the number of commuters for some working days, mandatory office days may observe pre-covid patterns. Further, wireless networks are dimensioned based on peak hour needs, which will coincide with mandatory office days.

Static urban densities

Table 4 shows some characteristics of the postal code understudy in comparison to postal code averages. This data is extracted from the open Paavo repository produced by Statistics Finland (Official Statistics of Finland, 2019a) except for the retail density, which is extracted from the "Helsinki Marketing's database of places" via the MyHelsinki Open API (City of Helsinki, 2020). We preprocessed data, selecting the most relevant features for device density calculation. For workplace density, we remove workplaces not belonging to the service industry (i.e., 6% of the total) since employees may work elsewhere than in the postal code.

For retailer density, we consider businesses in the field of shopping, accommodation, restaurants & cafes, museums & galleries, banquet venues, bars & nightlife, sauna & wellness since these typically use PoS devices. For building density, we include residential, commercial, office, and warehouse buildings. In our method implementation, we directly use retailer density and building density as static urban densities (see Table 2). The rest of the urban densities are derived from Table 4 values with additional Helsinki transportation data sources, as described in the next subsection.

*Table 4 Characteristics of the studied postal code*

|  | Studied postal code | Postal code average |
|---|---|---|
| Postal code | 00100 | - |
| Name | Helsinki Keskusta - Etu-Töölö | - |
| Area (sq. km) | 2.35 | 2.54 |
| Population density | 7,770 | 4,303 |
| Resident employed density | 4,262 | 2,129 |
| Workplace density | 19,785 | 3,403 |
| Retailer density | 197 | - |
| Building density | 269 | 226 |

Dynamic urban densities

To generate dynamic urban densities that change during 0-23h, we model the flow of people, passenger cars, buses, and bikes that enter and exit the postal code based on Helsinki transportation data on crossings through the border of the Helsinki peninsula. This peninsula consists of the 8 southernmost postal codes of Helsinki, including the one understudy. During the autumn of 2018, an average of 686,000 people crossed the peninsula border in either direction by car or public transport. Moreover, 31,400 people did it on a bicycle. We estimate the time of crossing for different transportation modes, assuming that they share similar time patterns based on statistics on motor vehicle traffic (City of Helsinki, 2019).

To estimate the active population density, we first calculate the number of people included in the peninsula between 0-23h due to crossings by subtracting those that leave from those that arrive and accumulating the sum. We normalize values between 0 and 1, generating what we call the peninsula-included people time pattern or *included people pattern* for short. Then, we estimate the active population density by adding the working population and the non-working population and dividing it by the postal code area. We model the working population to fluctuate during 0-23h according to the *included people pattern* with a maximum value equal to workplace density (as in Table 4). To calculate the non-working population, we assume that it equals the population density (as in Table 4) when the *included people pattern* is zero, and it evolves over time as people commute to work and back home. For example, in the morning hours, we make it decrease by gradually subtracting people until the total resident employed density (as in Table 5) is removed, which happens when the *included people pattern* is maximum.

Active, working, and non-working population densities describe population changes through the hours of the day, including both people on the move and static people. We generate a different density, called the moving population density, which only includes people on the move by subtracting active population density values between consecutive hours. The moving population density is used for calculating handover requests only. The densities presented next only include moving devices.

To estimate the moving car density, we first calculate the hourly number of cars in the postal code by adding/subtracting cars from the number of resident cars. We assume that the number of arriving/leaving cars during 0-14h/14-23h corresponds to the workforce deficit (i.e., workplaces minus employed residents), which fluctuate considering inward/outward crossing patterns observed on the peninsula border. We assume that cars account for 27% of the total commute volume (i.e., car transportation share), a car occupancy rate of 1.3, and an average car ownership per inhabitant of 0.329. Then, we estimate the moving car density as the difference in cars between consecutive hours (i.e., the number of cars that have arrived/left during that hour) divided by the area. In our calculation, the average car density equals 273 cars per square kilometer, which is a reasonable estimate compared to other cities, e.g., 325 in central London (Ajibulu et al., 2017).

We estimate moving bus density as 8% of the moving car density since buses account for 8% of the total passenger car count through the Helsinki peninsula (City of Helsinki, 2019b). We similarly estimate the moving bike density than for moving car density but also including the commute of residents employed in the same postal code. For this, we assume 6 bikes per 10 inhabitants and a bike transport share of 10% (City of Helsinki, 2019a; CONEBI, 2018). Figure 4 displays the obtained urban densities.

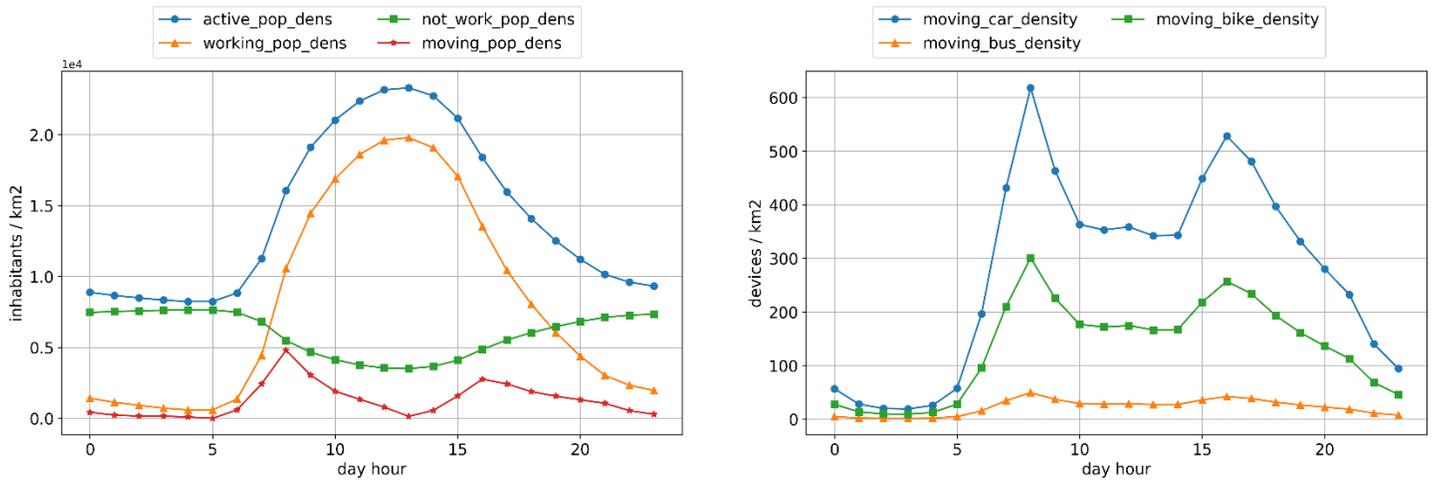

*Figure 4 Urban densities*

## 5. Results

### 5.1. Device penetration

Estimations based on the Bass model

To estimate the penetration of smartphones and mobile modems through the Finnish population, we fit a Bass model on historical GSMA data about the evolution of connections during 2007-2018 (GSMA, 2019). Then, we use the model to project their evolution. The obtained model parameters (p,q) are (0.036,0.016) and (0.021,0.089) for smartphone and mobile modem, respectively. While smartphones experience a small but continuous growth possibly driven by new business connections, the number of modems remains stable, which is coherent with the expectations that no additional fixed-to-mobile substitution will occur in Finnish households. We do not generate low/high scenarios since they already present high penetration over their potential user base, as noted in Section 3.5. Results on the device penetration are accumulated in Table 5.

Estimations based on replacement purchase models

We estimate the penetration of wearables through the Finnish population, estimating low/high penetrations to account for different consumer purchasing behaviors (i.e., use case-driven uncertainty). For the low growth scenario, we assume that connected smartwatches will remain as the dominant SIM-based wearable until 2030. In 2018, the first connected smartwatch was released, i.e., Apple iWatch version 3, followed by Android and Tizen-based. Nevertheless, the majority of current smartwatches rely on smartphones for communication. We assume that the smartwatch population's size increased during 2016-2018 from approximately 100,000 devices to 270,0000, as indicated by (AudienceProject, 2019). Also, smartwatch sales linearly increased following the trend observed during 2017-2018, when 166,000 and 178,000 were sold, respectively (Gotech et al., 2018). Moreover, we assume that SIM-based smartwatches represented 5% of the total sales in 2018, and this fraction doubles every year. Finally, the average smartwatch lifetime is 2 years. We generate a high penetration scenario by assuming that citizens will acquire more connected wearables in addition to smartwatches, including headphones, augmented reality glasses, connected shoes, etc. For this, we employ the same replacement purchase model with sales increasing twice as fast.

We estimate the penetration of connected cars through the population of passenger cars in use in the Uusimaa region[2]. This population has steadily increased during 2013-2018, with an annual average growth of 7,836 cars, reaching 710,457. The number of newly registered passenger cars has expanded with an average annual growth of approximately 50,000 (Official Statistics of Finland, 2019b). We assume that all newly registered passenger cars starting from 2018 onwards are equipped with mobile modems due to the EU mandate to support 112-based eCall. We estimate a low penetration scenario, considering 48,000 new annual registrations, population growth of 7,836, and a constant Uusimaa region population of 1,671,024. In contrast, we estimate a high penetration scenario motivated by transport electrification policies (e.g., subsidies on electric vehicle acquisition in exchange for polluting vehicles, road pricing for polluting vehicles), which may increase the number of newly registered passenger cars to 55,000 annually.

We estimate the penetration of connected bikes through the bikes population in use by employing the same method as for vehicles. Bikes are likely to get connected to enable tracking and thus discourage the stealing of electric bikes. A total of 12,000 were sold in Finland in 2018. We estimate a low penetration scenario, assuming a 1% increase in the total number of bikes after decommissions, 6 bikes in use every 10 inhabitants, bike sales to remain at similar 2016-2018 levels around 270,000 (Kuva, 2018), and a 50% annual increase in connected bike sales, which start at 5% of total bike sales from 2023. Alternatively, we estimate a high penetration scenario fueled by sustainable mobility policies (e.g., construction of bike lanes, parking spaces), which may increase bike sales to 310,000 annually, and a doubling number of connected bikes sales.

We estimate the penetration of autonomous buses through Helsinki public buses, i.e., approximately 1,500 vehicles. We estimate a low penetration scenario by assuming autonomous buses remain in a testing phase due to ineffective policy. For this, we speculate that every year a 12th of the stock is renewed (i.e., 121 buses, given that their lifecycle is about 12 years), and only 15% of newly acquired buses are autonomous, starting by 2022. For the high penetration scenario, we assume that more buses are annually acquired (i.e., 181 buses), and the share of autonomous buses doubles every year.

Estimations based on coverage requirements

We estimate street surveillance cameras' penetration through the postal code area, considering the restrictive/permissive approach from law enforcement policy. To the best of our knowledge, surveillance from existing street cameras is fragmented across owners (and locations), and the video is typically transported over wire. Deployment of open street video surveillance by the local police can be expected, given the commercial streets, tourist attractions, and government offices. For our estimation, we assume a Manhattan urban layout with blocks of 125 meters, leading to 64 intersections/km2. We generate a low and high penetration scenario by assuming 2 and 4 cameras per intersection, which implies a maximum of 128 cameras/km2 and 256 cameras/km2, respectively. Cameras are annually deployed, adding an additional 10% coverage starting in 2023. The number of cameras every 1,000 inhabitants by 2030 is 13 and 26 for the low and high penetration scenarios, respectively. In London, the most surveilled city in Europe, the average in 2018 was 68.4; 11.2 in Berlin (Biscchoff, 2019).

---

[2] The Uusima region contains the Greater Helsinki Area and surroundings. Vehicle registration data was accessible only at the region level.

We estimate urban sensors to reach a potential maximum of 1,629 per square kilometer, considering the connection of streetlights, air quality sensors, traffic lights, zebra crossing detectors, trash bins, bus stops, street ads, and infrastructure, such as bridges, tunnels, sewers. For this, we use the Manhattan urban layout defined for the open street video surveillance case, i.e., 64 blocks per square kilometer. We assumed a 50-meter distance between streetlights and trash bins leading to 9 items per intersection, 1 air quality sensor every four intersections, 4 traffic lights and zebra crossing detectors per intersection, 1 bus stop and 1 street advertisement per intersection (i.e., 24 sensors per intersection and 1,552 sensors per square kilometer, on average). Connected traffic lights and signs account for an additional 5% per square kilometer (i.e., 77 sensors per square kilometer). We estimate low and high penetration scenarios by assuming an annual sensor deployment speed equals 3% and 15% of the potential maximum, respectively, starting from 2023. The difference in deployment speeds is due to different policies on autonomous vehicles, air quality, and waste management.

The commercial utilization of autonomous drones for parcel delivery is limited by the availability of operator permits (given by the transport regulator) and landing areas, where shipments can be sent or dropped. We generate a low penetration scenario considering that parcels are only delivered between warehouses and drone-ready stores, e.g., urgent pharmaceutical prescriptions, starting by 2023. In this scenario, a maximum of 1 drone every 1,000 inhabitants is achieved over time, enabling 0.16 deliveries per inhabitant per week. For this, we assume that drones fly, on average, 15 km per delivery at 30 km/h, and the service is available 12 hours a day. We generate a high penetration scenario by assuming a wide availability of operator licenses and landing stations, enabling many delivery services, including food delivery. An average of 1.6 parcels per inhabitant is weekly delivered when the total drone population reaches its maximum of 1 drone every 100 inhabitants. In both cases, the number of drones increases annually, with 10% of the maximum number of drones.

Estimations via device-specific methods

We estimate the penetration of smart meters through the buildings in the postal code. The deployment of smart electricity meters has been completed in Finland through a regulatory mandate on providers. In 2014, a total of 3.4 million metering points were accounted for, with hourly metering and remote reading capabilities. However, smart water meters are not yet deployed, and testing is underway by some municipal water corporations (Digita, 2019). We generate a medium penetration forecast for water meters in which they are similarly deployed, like electricity meters. They are installed in 80% of households in 5 years, starting in 2022, following linear growth (Tractebel, 2019). We assume that buildings aggregate metering information for electricity and water via power line and local radio communications, which is afterward transmitted through two independent mobile connections. It is a convenient solution to retrofit old buildings that lack wired network access in the meter room. Smart meters deployed in Finland are equipped with several communication technologies, including 2G/3G communications, Zigbee and Mesh Radio Frequency technology (Landis+Gyr, 2010).

We estimate the penetration of POS devices through the retailers in the postal code. For the slow scenario, we assume that payment terminals remain as the dominant SIM-enabled retail device. The number of active payment terminals in Finland decreased during 2013-2015, possibly due to substitution by new cloud-based and smartphone-based payment systems (ECB, 2014). However, we estimate the penetration of payment terminals through city-center retailers to remain at 400% since many of them belong to well-established retail chains, which may prefer to use dedicated mobile payment terminals to guarantee direct communication with banks. We generate a high penetration scenario considering that, in addition to payment terminals, retailers adopt connected ads for shopfronts, consumer tracking solutions, and wireless bar-code readers for inventory management systems, etc. We estimate the number of connected devices per retailer to reach a maximum of 16 with an annual increase that equals 10% of the maximum number of devices, starting in 2021.

Results

*Table 5 Device penetration development*

| Device / year | Adopting body | Estimate | 2018 | 2019 | 2020 | 2021 | 2022 | 2023 | 2024 | 2025 | 2026 | 2027 | 2028 | 2029 | 2030 |
|---|---|---|---|---|---|---|---|---|---|---|---|---|---|---|---|
| Smartphones | Population | Medium | 1.48 | 1.52 | 1.55 | 1.57 | 1.59 | 1.61 | 1.62 | 1.63 | 1.64 | 1.65 | 1.66 | 1.66 | 1.66 |
| Mobile modems | Population | Medium | 0.30 | 0.30 | 0.30 | 0.30 | 0.31 | 0.31 | 0.31 | 0.31 | 0.31 | 0.31 | 0.31 | 0.31 | 0.31 |
| Wearables | Population | Low | 0.00 | 0.01 | 0.01 | 0.03 | 0.06 | 0.09 | 0.11 | 0.12 | 0.13 | 0.15 | 0.16 | 0.18 | 0.20 |
| Wearables | Population | High | 0.00 | 0.01 | 0.01 | 0.03 | 0.08 | 0.14 | 0.18 | 0.21 | 0.26 | 0.31 | 0.37 | 0.44 | 0.53 |
| Connected cars | Car stock | Low | 0.07 | 0.14 | 0.21 | 0.27 | 0.34 | 0.40 | 0.46 | 0.52 | 0.58 | 0.63 | 0.69 | 0.74 | 0.80 |
| Connected cars | Car stock | High | 0.08 | 0.15 | 0.23 | 0.30 | 0.37 | 0.44 | 0.51 | 0.57 | 0.64 | 0.70 | 0.77 | 0.83 | 0.89 |
| Autonomous buses | Public bus stock | Low | 0 | 0 | 0 | 0 | 0.01 | 0.03 | 0.04 | 0.05 | 0.06 | 0.08 | 0.09 | 0.10 | 0.11 |
| Autonomous buses | Public bus stock | High | 0 | 0 | 0 | 0 | 0.02 | 0.06 | 0.13 | 0.26 | 0.38 | 0.51 | 0.63 | 0.76 | 0.88 |
| Connected bikes | Bikes in use | Low | 0 | 0 | 0 | 0 | 0 | 0.01 | 0.01 | 0.02 | 0.04 | 0.06 | 0.08 | 0.13 | 0.19 |
| Connected bikes | Bikes in use | High | 0 | 0 | 0 | 0 | 0 | 9E-3 | 0.02 | 0.04 | 0.07 | 0.14 | 0.23 | 0.32 | 0.40 |
| Traffic lights, traffic signs, and urban sensors | Per km2 | Low | 0 | 0 | 0 | 0 | 0 | 49 | 98 | 147 | 195 | 244 | 293 | 342 | 391 |
| Traffic lights, traffic signs, and urban sensors | Per km2 | High | 0 | 0 | 0 | 0 | 0 | 325 | 651 | 977 | 1303 | 1629 | 1629 | 1629 | 1629 |
| Surveillance video cameras | Per km2 | Low | 0 | 0 | 0 | 0 | 0 | 13 | 26 | 38 | 51 | 64 | 77 | 90 | 102 |
| Surveillance video cameras | Per km2 | High | 0 | 0 | 0 | 0 | 0 | 26 | 51 | 77 | 102 | 128 | 154 | 179 | 205 |
| Autonomous drones | Population | Low | 0 | 0 | 0 | 1E-4 | 2E-4 | 3E-4 | 4E-4 | 5E-4 | 6E-4 | 7E-4 | 8E-4 | 9E-4 | 1E-3 |
| Autonomous drones | Population | High | 0 | 0 | 0 | 1E-3 | 2E-3 | 3E-3 | 4E-3 | 5E-3 | 6E-3 | 7E-3 | 8E-3 | 9E-3 | 1E-2 |
| Smart meters | Buildings | Medium | 1 | 1 | 1 | 1 | 1 | 1 | 2 | 2 | 2 | 2 | 2 | 2 | 2 |
| Retail devices | Retailers | Low | 4 | 4 | 4 | 4 | 4 | 4 | 4 | 4 | 4 | 4 | 4 | 4 | 4 |
| Retail devices | Retailers | High | 4 | 4 | 4 | 5 | 6 | 8 | 9 | 10 | 11 | 12 | 14 | 15 | 16 |

Note: 1E-3 stands for 1*10^(-3)

## 5.2. Device user volume

Estimation via exponential growth

We expect smartphone and modem traffic (which belong to the human traffic category, see Table 1) to grow during 2019-2030 due to increasing application bitrate requirements, e.g., caused by higher screen and camera resolutions. Also, due to the advent of new applications, e.g., smartphone-based virtual/augmented reality services[3]. In Finland, the annual traffic volume per broadband connection at the end of 2017 was approximately 98.7 GB, which equals 8.2 GB/month (Traficom, 2019). We estimate low/high growth based on Ericsson's country-level forecasts (Ericsson, 2019a, 2020). We generate a low estimate by assuming that volume increases at a 23% CAGR. This growth is moderately smaller than the expected average value for European countries during 2019-2025. For the high growth scenario, we assume that volume increases at a 28% CAGR, which is a slightly smaller growth than in countries where major service providers already launched 5G, such as the US and Middle East countries. We generate average daily volumes for the 2019-2030 period by uniformly distributing annual volumes to days. The estimated average daily volume for 2018 (i.e., 294 MB) roughly coincides with estimations from other studies based on crowd measurements (i.e., 287 MB). Next, we generate average hour-of-the-day volumes by distributing daily volumes according to crowdsourced device data from 22,795 users during July 2018 (Walelgne et al., 2019). Figure 5 shows the obtained results.

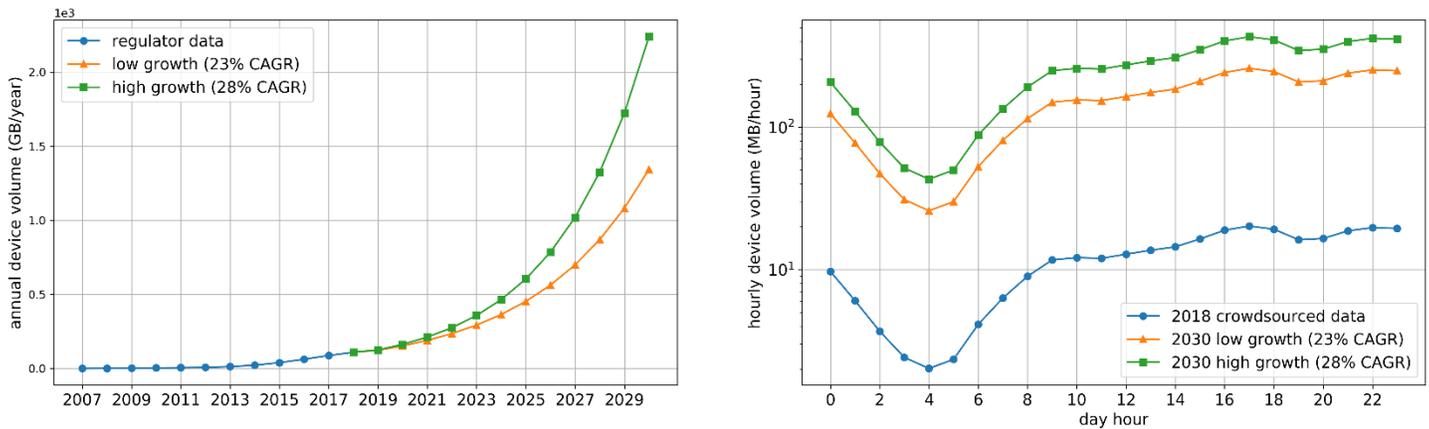

*Figure 5 Volume estimation for smartphones and modems*

Estimation via mixed exponential growth and ceiling-limited linear growth

We estimate daily traffic of connected cars, considering infotainment (which belongs to the human traffic category) and remote monitoring for technical services (which belongs to the machine high-activity traffic category). We estimate low/high infotainment volumes via exponential growth, following the same assumptions as for smartphones since they both are human traffic, and similar usage evolution can be expected. We assume an infotainment volume of 100 MB/day for 2019 since navigation and music streaming services typically consumed 100 MB/h, and the average driving time was 66 minutes (Traficom, 2019a). We assume that cars download operating system updates and inform about technical performance for vehicle technical services, consuming 72 MB/h on average. Since cars only transfer mobile data when moving and move for 66 minutes a day, the daily technical service volume is approximately 72 MB/day. We estimate technical service volumes via a ceiling-limited linear growth, given the advent of manufacturer-driven cellular vehicle-to-everything (C-V2X) services, for example, exchanging data about road driving conditions. However, we do not expect this traffic to increase in volume indefinitely but to stabilize by 2030 at approximately 362 MB/day and 723 MB/day (i.e., 5x/10x increase over 10 years) for the low and high growth estimates, respectively. For simplification, we assume the average daily driving time to remain constant at 1h. Figure 6 (left) shows the evolution of the daily device volume by aggregating application contributions and considering the total daily hours of device activity. In contrast, Figure 6 (right) indicates the hourly average volume of applications without considering device activity for 2030 only.

---

[3] To watch an e-streaming sports event in multi-view could consume about 7GB per hour, while a high-quality AR/VR stream with a median (bit) rate of 25Mbps would consume as much as 12GB per hour.

Estimation via ceiling-limited linear growth only

We estimate autonomous bus traffic generated by remote monitoring for technical services (which belongs to the machine high-activity traffic category) and by remote driving (that belongs to the high-priority traffic category). For technical services, we consider the same low/high growth estimations as connected cars but considering that buses drive for 10 hours every day, on average. For the remote driving application, we assume that buses are driven remotely for short periods, for example, during a technical malfunction. We generate a medium growth estimation that remains constant over the years. We do not expect the average remote driving time to increase over time since technology evolution should make remote driving better. Also, additional intervention time for new routes may be compensated by time reduction in old routes. We assume that autonomous buses, on average, require 0.5 hours/day of remote driving, which equals to 5 GB/day per bus. We equally distribute the total daily volume along the 10h of operation. We envision remote driving to require six HD video streams[4] (i.e., 2 for frontal vision, 2 for the rear sight, and 2 for in-bus vision) plus control instructions for steering and opening/closing doors.

We estimate delivery drone traffic following the same low/high growth estimations as for car remote monitoring. However, we consider that drones fly for 12 hours every day, on average, and the service is likely to start in 2021.

We expect surveillance cameras to process video locally and only to send part of this video over the network when additional processing or human intervention is needed, e.g., based on event-detection or citizen requests. We forecast video camera traffic, which belongs to the high-activity machine traffic category, by assuming that the amount of over-the-network video per camera annually increases due to growth in surveillance use cases and applications. In general, we assume that, in 2023, cameras start sending, on average, 7h of HD video footage every day. We estimate growth scenarios considering a permissive/restrictive data protection regulation affecting video data storage and processing. For the low growth estimation, the transmitted video annually increases by adding 1h/week, while for the high scenario, it increases by adding 4h/week. While the amount of video that humans can daily visualize is limited, video analytic resources can be increased to conduct tasks beyond the camera's local capabilities.

---

[4] We assume HD streams to have a resolution of 1920 x 1080 pixels (or 1080p) compressed via the H.264 standard, generating data streams of 10 Mbps.

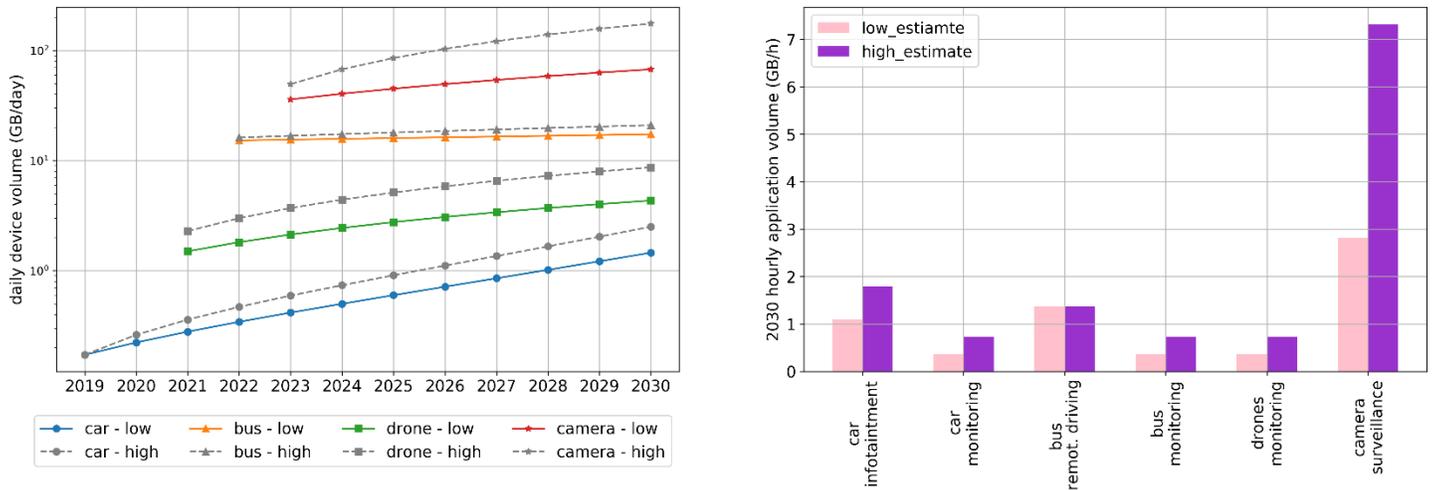

*Figure 6 Volume estimations for cars, buses, drones, and cameras*

### 5.3. Device control needs

MNO data suggests that IoT devices in Finland have a highly periodic activity with dominant intervals of 24h, 12h, 6h, and to a lesser extend, 1h (Finley et al., 2020). Based on this, we assume an average attachment inter-request time of 13h for 2019. Considering that new applications are more advanced and require equal or more frequent communication than their older versions, we estimate low/high control needs with 9h/7h inter-request time, respectively. Also, we suggest that the minimum inter-request time remains constant with a value of 15 minutes. Shorter values indicate that the device needs to be always connected, thus not behaving as a low-activity device. EU regulation on smart metering is planned to require real-time reporting every 15 min (Tractebel, 2019).

For the handover rate, we employ the previously mentioned Manhattan urban layout, i.e., an 8x8 street grid per km2. We assume that either vertical or horizontal street segments had, in 2019, 2 base station sites per MNO with an inter-site distance of 500m, implying a density of 16 sites/km2 per MNO. The total site density equals 48 sites/km2 that we consider a valid assumption since Finland has, by far, the highest site density of LTE eNodeBs in Europe (Rewheel, 2019). Considering that rapid demand will require a denser network, we estimate a low/high control needs with 400m/300m inter-site distance, respectively. We suggest the average speed of moving high-activity devices to remain at the same level, i.e., drones and cars at 30 km/h, buses at 20 km/h, smartphones at 15 km/h. Although applications for autonomous buses and drones could improve faster and increase device speed, their share of the device population is expected to remain small compared to connected cars.

### 5.4. Urban wireless traffic evolution

The daily user volume in the study area increases from 9 TB/km2 (in 2019) to 117 TB/km2 and 226 TB/km2 (in 2030), following a 27% and 34% CAGR growth for the slow and the rapid demand scenario, respectively. An increase in devices partially explains this growth. The median device density increases from 21,876 devices/km2 to 27,257 devices/km2 for the slow scenario (i.e., 2% CAGR) and to 36,112 devices/km2 for rapid scenario (i.e., 5% CAGR). These are average volumes and do no account for instantaneous peaks.

An overwhelming share of the traffic continues to be generated by smartphones. In 2019, they represented 89% of the traffic. In 2030, they respectively account for 80% and 69% for the slow and rapid scenarios. Connected cars and video cameras are the devices with the highest contribution to traffic after smartphones, with mobile modems contributing slightly less. In the slow demand scenario, connected cars generate 7% of the total traffic, while cameras a 6%. (i.e., 80% of cars are connected, and there are 13 cameras per 1,000 inhabitants). In the rapid demand scenario, cameras increase their contribution to 16%, while connected cars represent 7% (i.e., 89% of cars are connected, and there are 26 cameras per 1,000 inhabitants). Note that 7% of the slow scenario's daily volume means almost 8 TB/km2, while the corresponding number for the rapid is 15 TB/km2. The contribution of delivery drones and autonomous buses to the overall traffic remains below 1% across scenarios. The daily user volume evolution is presented in Figure 7, showing contribution per device and application.

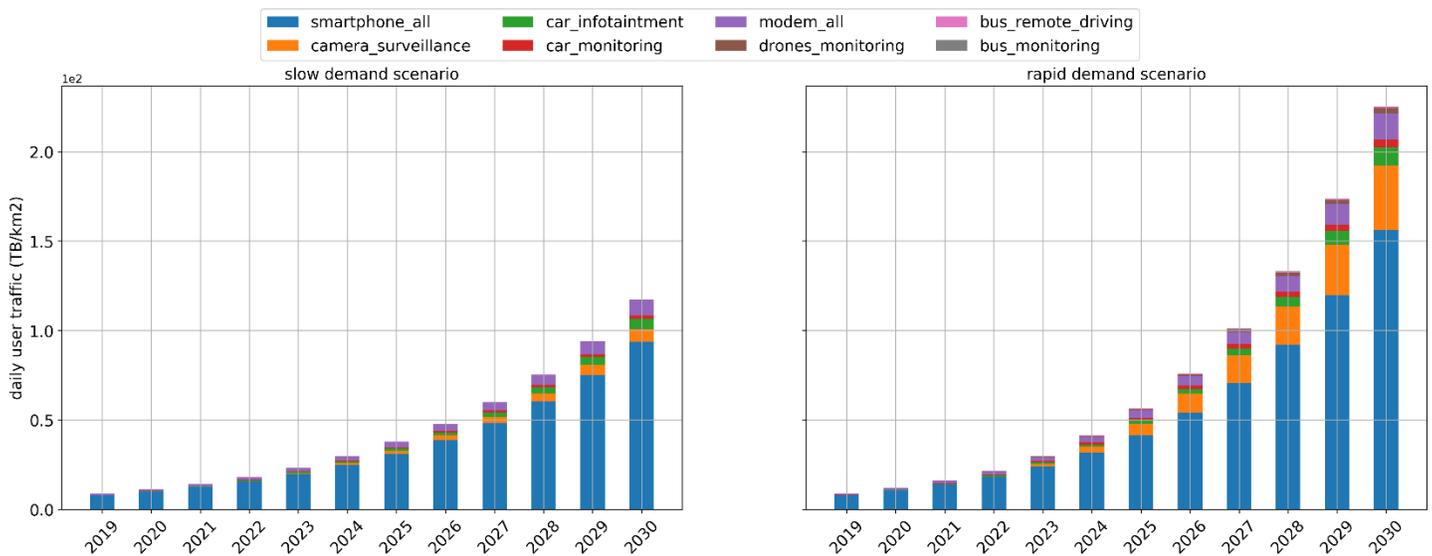

*Figure 7 Daily user volume evolution*

Evolution of traffic categories

A vast share of the traffic continues to belong to the human traffic category (i.e., from smartphones, modems, and car infotainment applications), which accounts for 92% and 80% of the total traffic for the slow and the rapid demand scenarios, respectively. In 2019, it represented 99% of the total traffic. In this traffic category, car infotainment traffic experiences the highest growth due to the faster adoption of connected cars than smartphones and modems. The daily median density of connected cars increases from 47 devices/km2 to 268 devices/km2 for the slow scenario, and to 301 devices/km2 for the rapid scenario. Compared to other categories, the total device density of this category experiences small growth, and its share decreases from 93% (in 2019) to 82% and 62% for the slow and fast demand scenarios, respectively.

High-activity machine traffic (i.e., from video cameras and car/bus/drone remote monitoring) experience different growth rates across scenarios, starting from 1% (in 2019) and achieving an 8% and a 19% share of the total daily traffic for the slow and the rapid demand scenarios. In the slow demand scenario, this traffic category grows at a 56% CAGR with volumes for drone and car remote monitoring as well as video cameras representing 2%, 21%, and 77% of the category share, respectively. In the rapid scenario, traffic grows at an 80% CAGR, with the mentioned shares changing to 7%, 10%, and 83%. For the slow demand scenario, the density of video cameras is 102 devices/km2 generating 7 TB/km2; for the high demand scenario, the density grows to 205 devices/km2 generating 36 TB/km2. Bus remote monitoring's traffic contribution remains small even for the rapid scenario (i.e., 0.3 TB/km2), given its low device density (24 devices/km2). In general, the daily median density of devices in this traffic category increases following 21% and a 28% CAGR for the slow and rapid scenarios, respectively. However, they only represent around 2% of the total device share. The daily user volume evolution is presented in Figure 8, showing contribution per device and application.

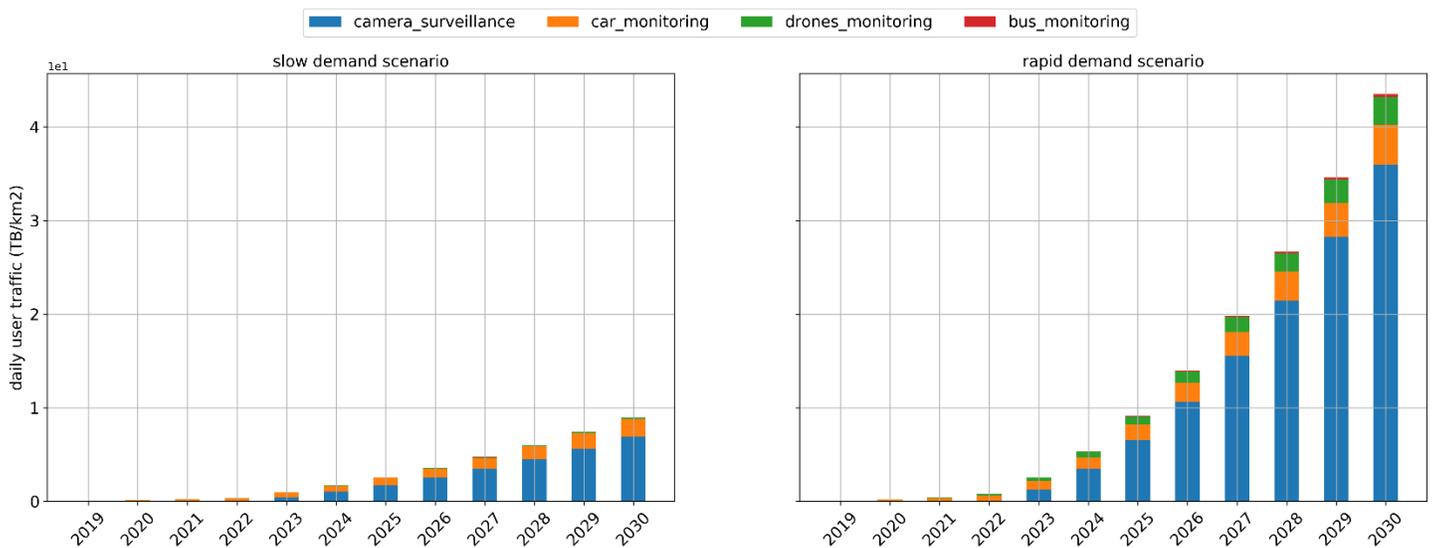

*Figure 8 Daily user volume evolution for high-activity machine traffic*

Low-activity machine traffic (i.e., from wearables, traffic lights and signs, connected bikes, urban sensors, smart meters, and POS) is irrelevant compared to other categories. In terms of daily median device density, they experience rapid growth, i.e., 11% CAGR for the slow scenario and 22% CAGR for the rapid scenario. Their share of the total device density radically increases from 6% to 16% and 36% for the slow and rapid demand scenario, respectively. The devices behind this growth are wearables, retail devices, urban sensors, smart meters, and connected bikes, in this order of importance. The difference between scenarios is caused by wearables, retail devices, and urban sensors in this order of importance.

High-priority traffic (i.e., from autonomous bus remote driving) observes different growth rates across scenarios, albeit it remains small compared to other traffic categories. For the low scenario, autonomous buses reach a median hour density of 3 devices/km2 producing 81 GB/km2. For the high scenario, the peak hour density is 24 devices/km2, generating 639 GB/km2.

Peak hour

The hour of the day with the highest traffic volume is the 16-17h. The peak hour volume increases from 0.7 TB/km2 to 9 TB/km2 and 16 TB/km2, following a 26% and a 33% CAGR growth for the slow and the rapid demand scenario. However, its traffic share slightly decreases from 7.7% (in 2019) to 7.5% and 7.2% (in 2030) for the mentioned scenarios. The hourly contribution of devices, including individual applications, is displayed in Figure 9.

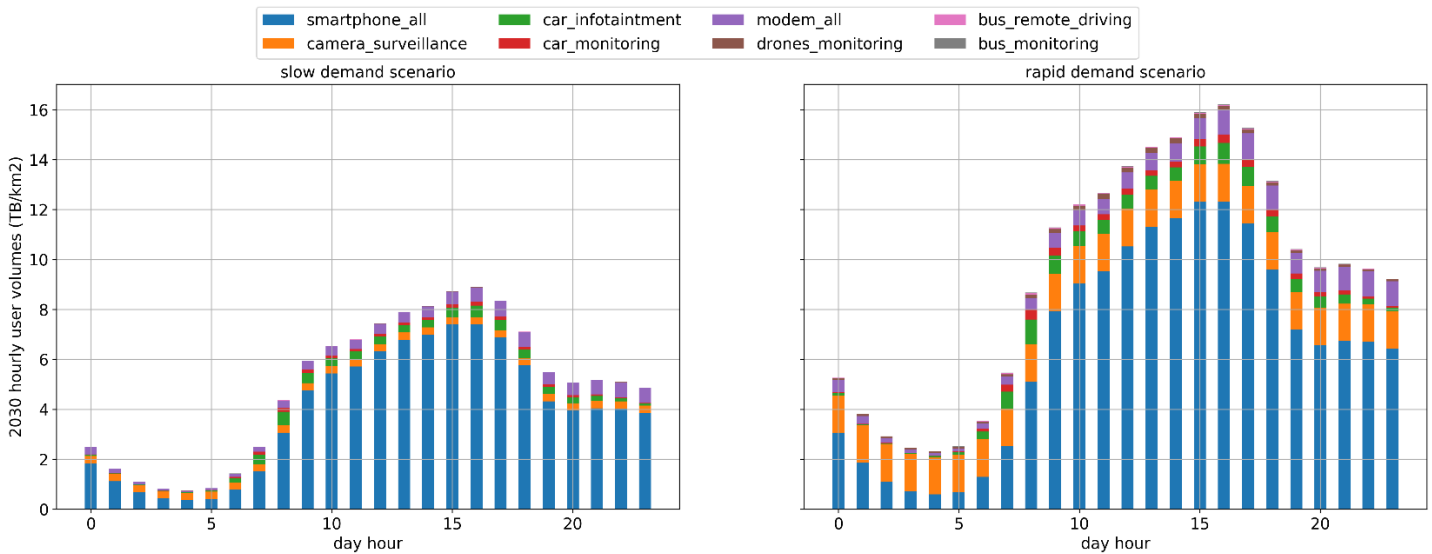

*Figure 9 Hourly user volumes in 2030*

The device density at the peak hour increases from 39,216 devices/km2 to 48,411 devices/km2 for the slow scenario (i.e., 2% CAGR) and to 61,205 devices/km2 for the rapid scenario (i.e., 4% CAGR). The share of peak hour devices vs. daily devices increases (concerning the daily median) from 5.5% (in 2019) to 5.6% and 5.9% (in 2030) for the mentioned scenarios.

In the peak hour of 2019, the network managed the attachment of smartphones and modems (i.e., 91% device share), payment terminals as well as smart meters (together 4%), and connected cars (below 1%). Hence, most of the intermittent attachment load was caused by low-activity devices (i.e., devices payment terminals and smart meters) since smartphones and modems were always active. Results indicate that the peak hour density of low-activity devices increases 16% CAGR and 30% CAGR for the slow and the rapid demand scenario, respectively. As a result, the attachment rate during the peak hour increases 6-fold and 21-fold for the mentioned scenarios, as shown in Figure 10 (left).

In the peak hour of 2019, handovers were mostly required by smartphones (90% device share) and some connected cars (below 1%). Results indicate that the peak hour density of moving high-activity devices increases following a 2% and 5% CAGR for the slow and the rapid demand scenarios, respectively. Without including smartphones, which are the ones with the slowest speed, the remaining devices (i.e., connected cars, delivery drones, and autonomous buses) observe a 20% and 28% CAGR increase for the mentioned scenarios. Therefore, the peak hour's handover rate increases 1.6-fold and 2.4-fold, as shown in Figure 10 (right).

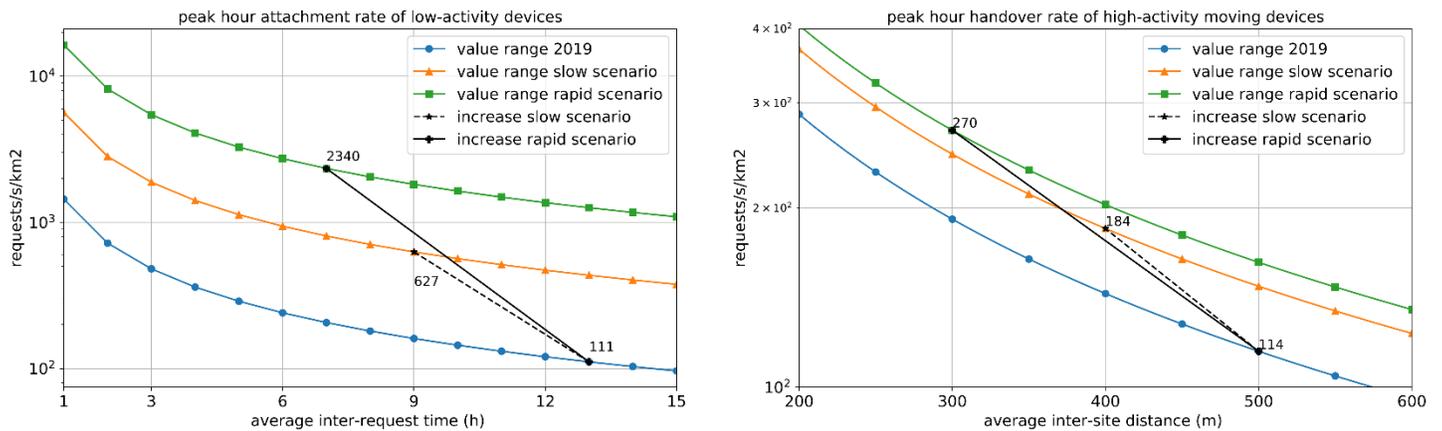

Figure 10 Network control needs

Detailed results on the 2019-2030 evolution of user volumes and device density, including individual applications and traffic categories, can be found in Annex B.

## 6. Discussion

### 6.1. Traffic demand sensitivity to multi-government policy

The effect of policy on device traffic depends on the device type. For example, policy effects are stronger for video cameras than for connected cars. Moreover, this effect should be understood considering two components, i.e., the effect on device penetration and the effect on device volume. Looking at device penetration, we observe that even when transport electrification policy increases car sales in a sustained manner, the diffusion of wireless technology into the car population rises only marginally. In contrast, when law enforcement policy embraces video surveillance, the deployment speed of cameras into an area is more rapid. Looking at device volume, we observe that video surveillance footage can be restricted to law enforcement purposes due to data protection regulation, halting the diffusion of application features. In that case, non-safety-related video is rarely sent over-the-network. In contrast, policy can hardly limit volumes for car remote monitoring or infotainment.

Policies seem to have a lesser effect on control traffic. Results indicate that wearables, PoS devices, and urban sensors are the primary sources of attachment requests at peak hour. The effect of policy on the penetration of the first two is small since private customer needs heavily drive purchasing. Only a subset of these devices could be affected by policy, including food tracking devices via safety food regulation. Smart city policies on air pollution control, waste management, and others can affect the deployment of urban sensors; however, these devices only represent between 1% and 3% of the peak hour density of low-activity devices. Looking at handovers, they continue to be driven by smartphones, followed by connected cars and delivery drones, and, to a lesser extent, autonomous buses. The effect of policy is most substantial for delivery drones since penetration depends on aviation regulation and landing areas' availability. Based on these observations, we suggest that the effect of policy is larger on devices providing public rather than private services.

### 6.2. Implications for operators

Pricing and revenue

Results indicate that daily volumes grow rapidly, primarily driven by human traffic, but also fueled by increasing machine traffic. Consequently, we suggest that in Finland, MNOs are likely to keep existing pricing structures for broadband devices, i.e., speed and volume-based monthly flat rates. MNOs may slightly modify these structures for machine high-activity traffic since they present broadband-like communication, except for traffic direction. Further, MNOs may add a premium on high-priority traffic, for example, based on session numbers and depending on time or location since complex network arrangements for packet prioritization and link redundancy are required. We latter discuss tradeoffs between net neutrality policies, aiming at equal traffic handling, and the prioritization of safety and mission-critical traffic.

Overall, MNO revenues likely remain constrained with potential small increases from machine high-activity traffic, e.g., via additions to existing consumer mobile broadband bills or enterprise contracts depending on the number of devices. Although high-priority traffic premiums could become lucrative, our forecast indicates slow adoption. Volume-based pricing structures can hardly yield revenues from machine low-activity traffic, given their shallow volumes.

New urban investment behavior

MNOs have managed to serve increasing broadband traffic by raising fees associated with subscription upgrades while investing in macro cell site upgrades and network automation. This investment behavior allowed increasing capacity while keeping costs under control since upgrades were typically limited to the addition of a new cell or antenna. However, this behavior may not be sustainable since macro cell-based capacity may eventually reach its limits, requiring the deployment of smaller cells. This limit is achieved as soon as spectrum from the 700 MHz and 3.5 GHz gets fully utilized. At this point, the 26 GHz band is needed, requiring street-level small cells, given its low-distance propagation characteristics (Busari et al., 2018). If the current investment behavior is followed, each MNO must build a new, very dense fiber network, including extensive urban civil works. In detail, small cells require a fiber fronthaul, i.e., fiber links are needed towards its first aggregation point, which need to be ducted below the city streets. Hence, cost decreasing trends for wireless capacity provision experience an inflection point since the cost of substantially adding capacity via small cells is significantly higher than via macro cells. Although cost evolution may reduce deployment cost in the long term, this reduction is limited to hardware equipment, and it can hardly affect civil works, which represent around 50% of the total cost depending on the deployment configuration (Landertshamer et al., 2019). Hence, to serve the estimated traffic volumes while preserving profit margins, MNOs are likely forced to mitigate costs by cooperating in the construction of a single, shared small cell network. In addition to cost, other bottlenecks favor this new urban investment behavior, such as limited site space, as discussed later in this section.

Technology and spectrum adoption options

To get a better grasp of when mmWave small cells can be needed, we discuss the timing for the Helsinki postal code under study. In 2019, the three Finnish MNOs had a combined spectrum of 240 MHz for download via 4G technology across the 700 MHz, 800 MHz, 1.8 GHz, 2.1 GHz, 2.6 GHz bands (Traficom, 2020). We suggest that, at best, MNOs could serve four times the 2019 peak hour volume just by updating existing macro cell sites. For this, they can exploit 390 MHz available for download at the 3.5 GHz band via 5G. In addition, they can increase the utilization of the upload spectrum since the 2030 peak hour volume is less skewed towards download than in 2019. Based on peak hour volume results, we suggest that mmWave small cell-based capacity may be required in the area under study around 2025. This requirement can be delayed until 2028 if massive multiple-input and multiple-output (mMIMO) technology gets broadly adopted or additional mid-band spectrum becomes available, e.g., on the 6,5 GHz band. Finally, MNO could also refarm 2.1 GHz and 2.6 GHz spectrum for 5G, additionally delaying small cell-based capacity needs until 2029[5].

Traffic forecast results indicate that not only user traffic increases, but also the control traffic. While small cells are well suited to serve a rising traffic demand, they may also increase the share of handover traffic due to shorter inter-site distances. Hence, technical optimizations for flexible handover may become necessary, that is networks that combine macro and small cells (Yamamoto et al., 2013). Results also indicate a significant increase in the rate of intermittent attachments by low-activity devices. In this case, the new Low Power Wide Area (LPWA) technologies already include numerous optimizations (e.g., LTE-M, 5G MIoT, LORA).

To a certain extent, this reasoning can be applied to other mid-sized European cities; however, deployment maybe later required outside Finland because the Finnish mobile network presents the highest base station site density in Europe (Rewheel, 2019).

### 6.3. Implications for telecom and local policy

The competitiveness of countries depends on their technological readiness and innovation capabilities among other factors (Sala-I-Martin et al., 2008; Schwab & Zahidi, 2020). Policymakers in the local and telecom context can improve technology readiness by making IoT connectivity promptly available in advanced urban areas, potentially enhancing firm productivity. Further, they can support technological innovation by enabling firms to pioneer the development and real-world testing of cutting-edge applications that require IoT connectivity (Caragliu & Del Bo, 2019).

---

[5] We don´t consider further macro cell densification as a long-term solution for capacity updates since Helsinki presents one of the highest base station site densities in Europe (Rewheel, 2019). The cost of new site deployment and its maintenance are likely to remain constant or even increase, as better sites (with easier access and better propagation characteristics) are already used. In addition to cost, increasing density of already crowded macro cell networks (as in Helsinki) may worse interference due to frequency reutilization. Hence, capacity based on mmWave bands maybe needed later in other European cities than in Helsinki.

IoT connection availability

IoT connectivity can be delayed when macro cell-based capacity gets fully utilized since revenue-constrained MNOs cannot afford to dig city streets, as previously discussed in this section. To address this first bottleneck, existing fiber networks for fixed access could be leveraged for mobile fronthaul. In this case, dependencies between fixed and mobile markets become stronger. While converged operators could swiftly deploy small cells, mobile-only operators face costly civil works. In other words, significant cost asymmetries arise between mobile market participants, depending on fixed access market shares. Markets with a dominant fixed operator may require access regulation that is specific for mobile fronthaul if no agreement can be reached with mobile-only operators. In markets with similarly sized converged operators, reciprocal access agreements might be likely for areas with no overlapping offering. However, access regulation may still be required to keep the remaining mobile-only operators in the game. Without intervention, dominant converged operators may slow down IoT connection availability, limiting deployment to a few hotspots. Multiple regulatory options exist to encourage fiber access for mobile fronthaul ranging from obligations to share (strong) to access pricing (soft) (Abrardi & Cambini, 2019; Bourreau et al., 2015).

Assuming that existing fiber networks are used for mobile fronthaul, small cell antennas still need to be located at certain height over the pavement and connected to the nearest fiber point of presence. A few sweet spots exist on the urban environment satisfying these requirements, including bus stops and lampposts. In contrast to the previous bottleneck, this second bottleneck is not driven by cost but by the lack of suitable sites. Challenges associated with limited space can be addressed via infrastructure sharing. Since this part of the network needs to be constructed from scratch, interested players may prefer to co-invest in a single, shared network to avoid individual MNO-municipality negotiations and bureaucratic overhead. Competition concerns on the monopoly of mmWave antennas should be addressed promoting service-based competition, potentially via 5G network slicing and neutral host business models. Mobile networks in underground transportation systems are a similar example (Lähteenmäki, 2021). Simultaneously, local policymakers may facilitate investments since lampposts and bus stops are already open for business, i.e., advertisement. Further, to harmonize deployment under municipal aesthetic guidance, mitigating uncoordinated installation of odd-looking hardware in touristic city center areas.

The third bottleneck is the mmWave spectrum which has been allocated in many European countries via country-wide[6] and local licenses. Despite country differences, substantial mmWave spectrum is available to MNOs, enabling investment.

---

[6] Examples of country-wide licenses in the 26 GHz band include Italy (1000 MHz unequally divided to 5 MNOs in 2018), Finland (2400 MHz equally divided to 3 MNOs in 2018), Greece (1000 MHz unequally divided to 3 MNOs in 2020), Denmark (2850 MHz unequally divided to 3 MNOs in 2021), and others.

26 GHz spectrum is also made available through different local licensing approaches, e.g., club use model in Italy, technical licenses in Finland. These local licenses typically allow for coverage that is limited to campuses or industrial parks, preventing non-MNO players to provide city-wide coverage, unless spectrum is leased from MNOs in the first place.

A small-scale example of urban small cell deployment is conducted by Freewave. This UK infrastructure company runs a 10-site outdoor small cell network in London's city center, implementing a neutral host business model (Freshwave, 2022). Generally, co-investment and ownership can be organized in multiple other ways, ranging from a joint MNO venture to a municipality-driven neutral operator. These configurations can vary across countries depending on spectrum and competition policy (Benseny et al., 2019; Benseny et al., 2020).

IoT-based application development

The development and testing of pioneer IoT applications can be delayed by unaware demand and net neutrality regulation.

Our forecast results suggest that the effect of policy on device traffic is larger for devices providing public rather than private services, as stated in Section 6.1. This assertion seems plausible considering that local policy can expand budgets of city subcontractors, who provide services with significant coverage needs, such as traffic signs and lights, surveillance cameras, urban sensors, etc. Hence, local policymakers can stimulate demand by pioneering the adoption of IoT applications for public service provision, in line with (Caragliu & Del Bo, 2019). For example, the diffusion of remote vehicle driving services could first emerge in public buses rather than on private cars because municipalities may have tighter control over the renewal of the rather small population of buses, i.e., device penetration effect. Moreover, local policymakers can modify both the daily bus routines and the urban environment (e.g., traffic signs, dedicated lanes), thus lowering the requirements for the diffusion of new autonomous driving applications.

Net neutrality policies, which aim for the equal handling of traffic, may inadvertently prevent the development of socially desirable applications covering public safety and mission-critical cases (Frias & Pérez Martínez, 2018). Countries aiming to adopt autonomous systems for transport (buses/trains) should introduce traffic prioritization exceptions in neutrality regulations. Operator investment may be encouraged if prioritized traffic can be priced differently.

In any case, policymakers should check for misalignments between policies across administrative levels. For autonomous buses, the urban environment's modification may not be sufficient if autonomous vehicle policies at the country-level do not allow driverless operations.

Summary

Table 6 summarizes bottlenecks for urban small cell deployment and its regulatory mitigation.

*Table 6 Bottlenecks and policy mitigation*

| Num. | Bottleneck | Scope | Government level | Mitigation |
|---|---|---|---|---|
| 1 | Access to fiber networks for mobile fronthaul | Supply | National | Access regulation for mobile fronthaul if agreement cannot be reached. |
| 2 | Limited street space for small cell sites, e.g., lampposts, bus stops. | Supply | National | To encourage infrastructure co-investment and sharing, monitoring service-based competition. |
| | | | Local | To facilitate coordinated deployment of antennas in lampposts, bus stops, etc. |
| 3 | mmWave spectrum | Supply | National | Continued allocation of high frequency bands. |
| 4 | Unaware demand | Demand | Local | To encourage IoT usage in public service provision. |
| 5 | Inability to prioritize traffic due to net neutrality regulation | Demand | National | To introduce traffic prioritization exceptions for socially desirable applications. |

### 6.4. Scientific implications

This article improves understanding of wireless traffic evolution by characterizing evolution through its fundamental underlying processes. The proposed method allows investigating traffic evolution considering a broader scope than before. It incorporates new inputs, such as the effect of policy, and it employs existing complementary methods to integrate a wide range of known inputs. The effect of policy is modeled in detail to affect individual device categories. Specifically, each category can be affected by a different policy combination, and the effect can be characterized through two components, i.e., the effect on device penetration and the effect on device user volume. Significantly, this study contributes explicitly to the understanding of traffic evolution for urban areas. Moreover, the provided analytical formulation can be re-used and extended to include an arbitrary number of devices, applications, traffic categories, and network technologies. Therefore, it can be used as a tool by operators and policymakers to anticipate the needs and consequences of network densification in the context of increasing traffic demand. The method provides a means to generate synthetic urban densities based on vehicle traffic flow and census data. Given the broad scope of the research method, implications for numerous ecosystem actors can be drawn.

### 7. Conclusions

This article reveals changes in urban wireless traffic derived from new IoT devices and policy effects from multiple government levels. Based on these changes, we provide guidance to operators for smooth mobile network evolution through mmWave small cell deployment. We also provide guidance to policymakers for IoT-enabled competitive gains via the mitigation of five bottlenecks.

We characterize the evolution of urban wireless traffic through the diffusion of new device features, the diffusion of new application features, and application usage development. Based on this new understanding, we propose a holistic method for investigating traffic evolution, widening the scope of previous research, given the advent of new devices and the growing role of local governments. We specifically consider the effect of policies in transport and mobility, law enforcement, digital economy, buildings, urban monitoring, and aviation on the traffic evolution of 11 device categories. We estimate traffic evolution for a postal code of the Helsinki city center during the 2020-2030 period.

Results indicate that daily user volumes grow fast, although a surprisingly large part of the traffic continues to be generated by smartphones. Machine traffic gains importance, reaching values amid 8% and 19%, mainly due to surveillance video cameras and connected cars. While camera traffic is sensitive to law enforcement policy and data regulation, car traffic is less affected by transport electrification policy. High-priority traffic remains small even when autonomous bus deployment is encouraged by autonomous vehicle policy. The control traffic at peak hour is likely to grow due to rising attachments from low-activity devices. Also, due to increasing handovers, albeit to a lower degree. In general, control traffic is less sensitive to policy than user traffic. We generally refer to average traffic volumes not accounting for instantaneous peaks.

Considering results, we observe that mobile networks need to transition from macro cell to heterogenous architectures to exploit mmWave frequencies via street-level small cells. We indicate that MNOs, who remain revenue-constrained, most likely need to cooperate in constructing a single, shared small cell network since this new infrastructure introduces an inflection point on the decreasing trend of wireless capacity provision cost.

Countries aiming to leverage wireless technologies for higher competitiveness should encourage investment in this infrastructure by facilitating access to existing fixed networks, allowing mmWave infrastructure monopoly in return for infrastructure sharing neutral models, and continuing the early allocation of high frequency bands. At the local level, city leaders should support coordinated deployment under municipal aesthetic guidance, continuing commercial exploitation of urban furniture, such as lampposts, bus stops. Wireless technologies can also increase competitiveness when employed for cutting-edge applications. Local policymakers can stimulate application development by pioneering IoT adoption in public services and introducing changes to the physical environment when needed. At the national level, exceptions to net neutrality regulation should be introduced to allow traffic prioritization for socially desirable applications.

Future research can target narrower urban areas, increasing the geographical resolution. Also, considering the development of urban densities, e.g., changes in transportation mode. The article observations primarily apply to busy urban areas, e.g., the central business district, of mid-sized cities in countries with advanced telecom markets.

**Acknowledgments**

This work was done as part of the Neutral Host Pilot project within the LuxTurrim5G ecosystem funded by the participating companies and Business Finland (https://www.luxturrim5g.com/).

# References

Abrardi, L., & Cambini, C. (2019). Ultra-fast broadband investment and adoption: A survey. *Telecommunications Policy*, *43*(3), 183–198. https://doi.org/10.1016/J.TELPOL.2019.02.005

Ajibulu, A., Bradford, J., Chambers, C., Karousos, A., & Konstantinou, K. (2017). *H2020 5G-NORMA Deliverable 2.3 - Evaluation architecture design and socio- economic analysis - final report*.

AudienceProject. (2019). *INSIGHTS 2018 - Device usage report*.

Bao, Y., Eddine El Ayoubi, S., Pujol, F., Manero, C., Copigneaux, B., Hossein, I., Widaa, A., Markendahl, J., & Zimmermann, G. (2017). *Deliverable D1.2 Quantitative techno-economic feasibility assessment Quantitative techno-economic feasibility assessment*.

Batty, M., Axhausen, K. W., Giannotti, F., Pozdnoukhov, A., Bazzani, A., Wachowicz, M., Ouzounis, G., & Portugali, Y. (2012). Smart cities of the future. *Eur. Phys. J. Special Topics*, *214*, 481–518. https://doi.org/10.1140/epjst/e2012-01703-3

Bauer, J. M., & Bohlin, E. (2022). Regulation and innovation in 5G markets. *Telecommunications Policy*, *46*(4), 102260. https://doi.org/10.1016/J.TELPOL.2021.102260

Benseny, J., Walia, J., & Hämmäinen, H. (2020). Value Network Configurations for Smart Cities. *Journal of The Institute of Telecommunications Professionals*, *14*, 18–25.

Biscchoff, P. (2019). *Surveillance Camera Statistics: Which City has the Most CCTV Cameras?* https://www.comparitech.com/vpn-privacy/the-worlds-most-surveilled-cities/

Blind, K., & Niebel, C. (2022). 5G roll-out failures addressed by innovation policies in the EU. *Technological Forecasting and Social Change*, *180*, 121673. https://doi.org/10.1016/J.TECHFORE.2022.121673

Bourreau, M., Cambini, C., & Hoernig, S. (2015). Geographic access markets and investments. *Information Economics and Policy*, *31*, 13–21. https://doi.org/10.1016/J.INFOECOPOL.2015.04.003

Busari, S. A., Mumtaz, S., Al-Rubaye, S., & Rodriguez, J. (2018). 5G millimeter-wave mobile broadband: Performance and challenges. *IEEE Communications Magazine*, *56*(6), 137–143.

Caragliu, A., & Del Bo, C. F. (2019). Smart innovative cities: The impact of Smart City policies on urban innovation. *Technological Forecasting and Social Change*, *142*, 373–383. https://doi.org/10.1016/j.techfore.2018.07.022

Cici, B., Gjoka, M., Markopoulou, A., & Butts, C. T. (2015). On the decomposition of cell phone activity patterns and their connection with urban ecology. *Proceedings of the International Symposium on Mobile Ad Hoc Networking and Computing (MobiHoc)*, *2015-June*, 317–326. https://doi.org/10.1145/2746285.2746292

Cisco. (2020). *Cisco Annual Internet Report*.

[dataset] City of Helsinki, 2019a, Tilastotietoja Helsingin autoliikenteestä (Motor vehicle traffic volumes), version 13/11/2019, 2019, https://www.hel.fi/static/liitteet/kaupunkiymparisto/liikenne-ja-kartat/kadut/liikennetilastot/autoliikenne/hki_autoliikenteen_tilastot.xlsx

City of Helsinki. (2019b). *Motor vehicle traffic volumes*.

CONEBI. (2018). European Bicycle Market 2017 edition - Industry & MArket Profile. In *Associatioin of the European Two-Wheeler Parts' & Accessories' Industry*.

Digita. (2019). *Digita's IoT network is digitalising the water supply - Digita*. https://www.digita.fi/en/media/digita_s_iot_network_is_digitalising_the_water_supply.5678.news

Ding, J., Xu, R., Li, Y., Hui, P., & Jin, D. (2018). Measurement-Driven Modeling for Connection Density and Traffic Distribution in Large-Scale Urban Mobile Networks. *IEEE Transactions on Mobile Computing*, *17*(5), 1105–1118. https://doi.org/10.1109/TMC.2017.2752159

ECB. (2014). *Payments statistics : methodological notes* (Issue November 2007).

Ericsson. (2019a). *Ericsson Mobility Report June 2019*.

Ericsson. (2019b). *Ericsson Mobility Report November 2019*.

Ericsson. (2020). *Ericsson Mobility Report June 2020*.

Finley, B., Benseny, J., Vesselkov, A., & Walia, J. (2020). *How does IoT traffic evolve ? Real-world evidence from a European operator*. 1–24.

Fisher, J. C., & Pry, R. H. (1971). A simple substitution model of technological change. *Technological Forecasting and Social Change*, *3*(C), 75–88. https://doi.org/10.1016/S0040-1625(71)80005-7

Freshwave. (2022, December 22). *EE live in first-of-its-kind mobile infrastructure pilot in City of London with Freshwave*. https://freshwavegroup.com/ee-live-in-first-of-its-kind-mobile-infrastructure-pilot-in-city-of-london-with-freshwave/

Frias, Z., & Pérez Martínez, J. (2018). 5G networks: Will technology and policy collide? *Telecommunications Policy*, *42*(8), 612–621. https://doi.org/10.1016/J.TELPOL.2017.06.003

Gotech, GfK, & Tukkukauppiaat Elektroniikan. (2018). *Retail sales volume of mobile phones, smartphones and smartwatches in Finland*.

Grant-Muller, S., & Usher, M. (2014). Intelligent Transport Systems: The propensity for environmental and economic benefits. *Technological Forecasting and Social Change*, *82*(1), 149–166. https://doi.org/10.1016/j.techfore.2013.06.010

[dataset] GSMA, 2019. GSMA Intelligence Markets Data, version 05/10/2018, 2019, https://www.gsmaintelligence.com/data/

GSMA. (2020). *Lessons to accelerate economic growth and recovery Mobile technology and economic growth*. www.gsmaintelligence.com

Halepovic, E., Williamson, C., & Ghaderi, M. (2009). Wireless data traffic: A decade of change. In *IEEE Network* (Vol. 23, Issue 2, pp. 20–26). https://doi.org/10.1109/MNET.2009.4804332

Hatuka, T., & Zur, H. (2020). From smart cities to smart social urbanism: A framework for shaping the socio-technological ecosystems in cities. *Telematics and Informatics*, *55*, 101430. https://doi.org/10.1016/J.TELE.2020.101430

Jha, A., & Saha, D. (2020). "Forecasting and analysing the characteristics of 3G and 4G mobile broadband diffusion in India: A comparative evaluation of Bass, Norton-Bass, Gompertz, and logistic growth models." *Technological Forecasting and Social Change*, *152*, 119885. https://doi.org/10.1016/j.techfore.2019.119885

Jin, Y., Duffield, N., Gerber, A., Haffner, P., Hsu, W.-L., Jacobson, G., Sen, S., Venkataraman, S., & Zhang, Z.-L. (2012). Characterizing data usage patterns in a large cellular network. *Proceedings of the 2012 ACM SIGCOMM Workshop on Cellular Networks: Operations, Challenges, and Future Design*, 7–12.

Khan, Z., Anjum, A., Soomro, K., & Tahir, M. A. (2015). Towards cloud based big data analytics for smart future cities. *Journal of Cloud Computing*, *4*(1), 2. https://doi.org/10.1186/s13677-015-0026-8

Kivi, A., Smura, T., & Töyli, J. (2012). Technology product evolution and the diffusion of new product features. *Technological Forecasting and Social Change*, *79*(1), 107–126. https://doi.org/10.1016/j.techfore.2011.06.001

Knieps, G., & Bauer, J. M. (2022). Internet of things and the economics of 5G-based local industrial networks. *Telecommunications Policy*, *46*(4), 102261. https://doi.org/10.1016/J.TELPOL.2021.102261

Kuva, H. (2018). *Hot Summer Accelerated Bike Sales in Finland - Bike Europe*. Bike Europe. https://www.bike-eu.com/sales-trends/nieuws/2018/12/hot-summer-accelerated-bike-sales-in-finland-10134975

Kyriazis, D., Varvarigou, T., White, D., Rossi, A., & Cooper, J. (2013). Sustainable smart city IoT applications: Heat and electricity management & Eco-conscious cruise control for public transportation. *2013 IEEE 14th International Symposium on a World of Wireless, Mobile and Multimedia Networks, WoWMoM 2013*. https://doi.org/10.1109/WoWMoM.2013.6583500

Lähteenmäki, J. (2021). The evolution paths of neutral host businesses: Antecedents, strategies, and business models. *Telecommunications Policy*, *45*(10), 102201. https://doi.org/10.1016/J.TELPOL.2021.102201

Landertshamer, O., Benseny, J., Hämmainen, H., & Wainio, P. (2019). Cost model for a 5G smart light pole network. *2019 CTTE-FITCE: Smart Cities and Information and Communication Technology, CTTE-FITCE 2019*. https://doi.org/10.1109/CTTE-FITCE.2019.8894818

Landis+Gyr. (2010). *Landis+Gyr agreement in Helsinki is at the forefront of smart meter deployment in EU - Landis+Gyr*. https://www.landisgyr.eu/news/landisgyr-agreement-in-helsinki-is-at-the-forefront-of-smart-meter-deployment-in-eu/

Lee, S., Cho, C., Hong, E. K., & Yoon, B. (2016). Forecasting mobile broadband traffic: Application of scenario analysis and Delphi method. *Expert Systems with Applications*, *44*, 126–137. https://doi.org/10.1016/j.eswa.2015.09.030

Lehr, W., Queder, F., & Haucap, J. (2021). 5G: A new future for Mobile Network Operators, or not? *Telecommunications Policy*, *45*(3), 102086. https://doi.org/10.1016/j.telpol.2020.102086

Lim, J., Nam, C., Kim, S., Rhee, H., Lee, E., & Lee, H. (2012). Forecasting 3G mobile subscription in China: A study based on stochastic frontier analysis and a Bass diffusion model. *Telecommunications Policy*, *36*(10–11), 858–871. https://doi.org/10.1016/J.TELPOL.2012.07.016

Lin, T., Rivano, H., & Le Mouel, F. (2017). A Survey of Smart Parking Solutions. *IEEE Transactions on Intelligent Transportation Systems*, *18*(12), 3229–3253. https://doi.org/10.1109/TITS.2017.2685143

Liu, L., Essam, D., & Lynar, T. (2022). Complexity Measures for IoT Network Traffic. *IEEE Internet of Things Journal*, *9*(24), 25715–25735. https://doi.org/10.1109/JIOT.2022.3197323

Liu, X., Li, H., Lu, X., Xie, T., Mei, Q., Feng, F., & Mei, H. (2018). Understanding Diverse Usage Patterns from Large-Scale Appstore-Service Profiles. *IEEE Transactions on Software Engineering*, *44*(4), 384–411. https://doi.org/10.1109/TSE.2017.2685387

Maeng, K., Kim, J., & Shin, J. (2020). Demand forecasting for the 5G service market considering consumer preference and purchase delay behavior. *Telematics and Informatics*, *47*, 101327. https://doi.org/10.1016/J.TELE.2019.101327

Malandrino, F., Chiasserini, C. F., & Kirkpatrick, S. (2017). Cellular network traces towards 5G: Usage, analysis and generation. *IEEE Transactions on Mobile Computing*, *17*(3), 529–542. https://doi.org/10.1109/TMC.2017.2737011

Meade, N., & Islam, T. (2006). Modelling and forecasting the diffusion of innovation - A 25-year review. *International Journal of Forecasting*, *22*(3), 519–545. https://doi.org/10.1016/j.ijforecast.2006.01.005

Nam, C., Kim, S., & Lee, H. (2008). The role of WiBro: Filling the gaps in mobile broadband technologies. *Technological Forecasting and Social Change*, *75*(3), 438–448. https://doi.org/10.1016/J.TECHFORE.2007.04.008

Neirotti, P., De Marco, A., Cagliano, A. C., Mangano, G., & Scorrano, F. (2014). Current trends in smart city initiatives: Some stylised facts. *Cities*, *38*, 25–36. https://doi.org/10.1016/j.cities.2013.12.010

OECD. (2017a). OECD Digital Economy Outlook 2017. In *OECD Digital Economy Outlook 2017*. OECD. https://doi.org/10.1787/9789264276284-en

OECD. (2017b). *OECD Science, Technology and Industry Scoreboard 2017*. OECD. https://doi.org/10.1787/9789264268821-en

[dataset] Official Statistics of Finland, 2019b, Paavo Open data by postal code area, version 2019, 2019, http://www.stat.fi/tup/paavo/index_en.html

Official Statistics of Finland. (2019b). *Motor vehicle stock*. https://www.stat.fi/til/mkan/tau_en.html

Olson, J., & Choi, S. (1985). A product diffusion model incorporating repeat purchases. *Technological Forecasting and Social Change*, *27*(4), 385–397. https://doi.org/10.1016/0040-1625(85)90019-8

Rathore, M. M., Ahmad, A., Paul, A., & Rho, S. (2016). Urban planning and building smart cities based on the Internet of Things using Big Data analytics. *Computer Networks*, *101*, 63–80. https://doi.org/10.1016/J.COMNET.2015.12.023

Rewheel. (2019). *Site density is key to LTE network performance – and critical for 5G*.

Riikonen, A., Smura, T., Kivi, A., & Töyli, J. (2013). Diffusion of mobile handset features: Analysis of turning points and stages. *Telecommunications Policy*, *37*(6–7), 563–572. https://doi.org/10.1016/j.telpol.2012.07.011

Romirer-Maierhofer, P., Schiavone, M., & D'Alconzo, A. (2015). Device-specific traffic characterization for root cause analysis in cellular networks. *Lecture Notes in Computer Science (Including Subseries Lecture Notes in Artificial Intelligence and Lecture Notes in Bioinformatics)*, *9053*, 64–78. https://doi.org/10.1007/978-3-319-17172-2_5

Sala-I-Martin, X., Blanke, J., Drzeniek, M., Thierry Geiger, H., Mia, I., & Paua, F. (2008). *The Global Competitiveness Index: Prioritizing the Economic Policy Agenda*.

Schneir, J. R., Ajibulu, A., Konstantinou, K., Bradford, J., Zimmermann, G., Droste, H., & Canto, R. (2019). A business case for 5G mobile broadband in a dense urban area. *Telecommunications Policy*.

Schwab, K., & Zahidi, S. (2020). *The Global Competitiveness Report How Countries are Performing on the Road to Recovery*. www.weforum.org

Shafiq, M., Tian, Z., Bashir, A. K., Jolfaei, A., & Yu, X. (2020). Data mining and machine learning methods for sustainable smart cities traffic classification: A survey. *Sustainable Cities and Society*, *60*, 102177. https://doi.org/10.1016/J.SCS.2020.102177

Shafiq, M. Z., Ji, L., Liu, A. X., Pang, J., & Wang, J. (2013). Large-scale measurement and characterization of cellular machine-to-machine traffic. *IEEE/ACM Transactions on Networking*, *21*(6), 1960–1973. https://doi.org/10.1109/TNET.2013.2256431

Tahaei, H., Afifi, F., Asemi, A., Zaki, F., & Anuar, N. B. (2020). The rise of traffic classification in IoT networks: A survey. *Journal of Network and Computer Applications*, *154*, 102538. https://doi.org/10.1016/J.JNCA.2020.102538

[dataset] The city of Helsinki. (2020), MyHelsinki Open API, version unknown, 2020, open-api.myhelsinki.fi/

Tractebel. (2019). European Smart Metering Benchmark. *European Commission DG Energy - Belgium*, *June*, 1–129. https://doi.org/10.5771/9783845266190-974

[dataset] Traficom, 2019a. Table on communications service statistics, version 30/06/2019, 2019, https://www.traficom.fi/sites/default/files/media/publication/Viestintapalveluiden-tilastokoonti.ods

Traficom. (2019b). *Volume of data transferred in mobile network*. https://www.traficom.fi/en/statistics/volume-data-transferred-mobile-network

Traficom. (2020). *Frequencies and license holders of public mobile networks*. https://www.traficom.fi/en/communications/communications-networks/frequencies-and-license-holders-public-mobile-networks

Walelgne, E. A., Asrese, A. S., Manner, J., Bajpai, V., & Ott, J. (n.d.). *Understanding Data Usage Patterns of Geographically Diverse Mobile Users*.

Walelgne, E. A., Asrese, A. S., Manner, J., Bajpai, V., & Ott, J. (2021). Understanding Data Usage Patterns of Geographically Diverse Mobile Users. *IEEE Transactions on Network and Service Management*, *18*(3), 3798–3812. https://doi.org/10.1109/TNSM.2020.3037503

Xu, F., Li, Y., Wang, H., Zhang, P., & Jin, D. (2017). Understanding Mobile Traffic Patterns of Large Scale Cellular Towers in Urban Environment. *IEEE/ACM Transactions on Networking*, *25*(2), 1147–1161. https://doi.org/10.1109/TNET.2016.2623950

Yleisradio. (2019). *Drone-assisted food deliveries to begin in Helsinki | Yle Uutiset | yle.fi*. https://yle.fi/uutiset/osasto/news/drone-assisted_food_deliveries_to_begin_in_helsinki/10789225

Ylipulli, J., & Luusua, A. (2020). Smart cities with a Nordic twist? Public sector digitalization in Finnish data-rich cities. *Telematics and Informatics*, *55*, 101457. https://doi.org/10.1016/J.TELE.2020.101457

## A. Annex A - Traffic categories

We propose four traffic categories, as presented in Table A.1. These categories result from the study of identified devices and applications, considering their dominant source of traffic (i.e., Human, Machine, Event), the level of traffic generating activity (i.e., High, Low), and the traffic priority (i.e., Low, Medium, High), as described in Table A.2.

*Table A.1 Traffic categories*

| Number | Name | Description |
| --- | --- | --- |
| 1 | Human traffic | Human-generated high-activity medium-priority traffic |
| 2 | Machine low-activity traffic | Machine-generated low-activity low-priority traffic (also known as wide area IoT) |
| 3 | Machine high-activity traffic | Machine-generated high-activity medium-priority traffic (also known as short range IoT) |
| 4 | High-priority traffic | Event-based high-priority traffic |

*Table A.2 Analysis of device traffic characteristics*

| Devices | Application | Dominant source | Traffic activity | Traffic priority | Traffic category |
| --- | --- | --- | --- | --- | --- |
| Smartphones and mobile modems | Multiple via app. market | Human usage | High | Medium | (1) Human traffic |
| Wearables | Multiple via app. market | Machine monitoring | Low | Low | (2) Machine low-activity traffic |
| Connected bikes | Tracking | Machine monitoring | Low | Low | 2 |
| Connected vehicles | Remote monitoring | Machine monitoring | High | Medium | (3) Machine high-activity traffic |
| Connected vehicles | Multiple via app. market (Infotainment) | Human usage | High | Medium | 1 |
| Traffic signs and lights | Sensing and acting | Machine monitoring | Low | Medium | 2 |
| Autonomous buses | Remote driving | Event-based | High | High | (4) High-priority traffic |
| Autonomous buses | Remote monitoring | Machine monitoring | High | Medium | 3 |
| Street surveillance video cameras | Video streaming and processing | Machine monitoring | High | Medium | 3 |
| Urban sensors | Sensing and acting | Machine monitoring | Low | Low | 2 |
| Electricity and water meters | Sensing and acting | Machine monitoring | Low | Low | 2 |
| Parcel delivery drones | Remote monitoring | Machine monitoring | High | Medium | 3 |
| Point of sale (POS) devices | Payment, logistics | Event-based | High | Medium | 2 |

## B. Annex B - Detailed results

In this section, we share detailed results on user traffic evolution and device density evolution between 2019 and 2030, including slow and rapid demand scenarios.

*Table B.1 Daily user volume evolution*

|  | 2019 | | 2030 - slow demand scenario | | | 2030 - rapid demand scenario | | |
| --- | --- | --- | --- | --- | --- | --- | --- | --- |
| Traffic category and device | density [GB/km2] | share | density [GB/km2] | share | CAGR | density [GB/km2] | share | CAGR |
| **High-priority traffic** | **0** | **0 %** | **81** | **0 %** | **49 %** | **639** | **0 %** | **80 %** |
| bus_remote_driving | 0 | 0 % | 81 | 0 % | 49 % | 639 | 0 % | 80 % |
| **Human traffic** | **8,673** | **99 %** | **108,373** | **92 %** | **26 %** | **181,416** | **80 %** | **32 %** |
| smartphone_all | 7,793 | 89 % | 93,807 | 80 % | 25 % | 156,217 | 69 % | 31 % |
| modem_all | 787 | 9 % | 8,795 | 7 % | 25 % | 14,647 | 6 % | 30 % |
| car_infotainment | 93 | 1 % | 5,771 | 5 % | 46 % | 10,552 | 5 % | 54 % |
| **Machine high-activity traffic** | **67** | **1 %** | **8,986** | **8 %** | **56 %** | **43,518** | **19 %** | **80 %** |
| car_monitoring | 67 | 1 % | 1,905 | 2 % | 36 % | 4,277 | 2 % | 46 % |
| camera_surveillance | 0 | 0 % | 6,912 | 6 % | 123 % | 35,942 | 16 % | 160 % |
| drone_monitoring | 0 | 0 % | 148 | 0 % | 58 % | 2,962 | 1 % | 107 % |
| bus_monitoring | 0 | 0 % | 21 | 0 % | 32 % | 337 | 0 % | 70 % |
| **Total** | **8,740** | **100 %** | **117,440** | **100 %** | **27 %** | **225,573** | **100 %** | **34 %** |

*Table B.2 Peak hour user volume evolution*

|  | 2019 | | 2030 - slow demand scenario | | | 2030 - rapid demand scenario | | |
| --- | --- | --- | --- | --- | --- | --- | --- | --- |
| Traffic category and device | density [GB/km2] | share | density [GB/km2] | share | CAGR | density [GB/km2] | share | CAGR |
| **High-priority traffic** | **0** | **0 %** | **6** | **0 %** | **0** | **51** | **0 %** | **43 %** |
| bus_remote_driving | 0 | 0 % | 6 | 0 % | 18 % | 51 | 0 % | 43 % |
| **Human traffic** | **673** | **100 %** | **8,441** | **94 %** | **96 %** | **14,133** | **87 %** | **116 %** |
| smartphone_all | 615 | 91 % | 7,403 | 83 % | 25 % | 12,328 | 76 % | 31 % |
| modem_all | 51 | 8 % | 577 | 6 % | 25 % | 962 | 6 % | 30 % |
| car_infotainment | 7 | 1 % | 461 | 5 % | 46 % | 843 | 5 % | 55 % |
| **Machine high-activity traffic** | **5** | **1 %** | **448** | **5 %** | **122 %** | **2,023** | **12 %** | **234 %** |
| car_monitoring | 5 | 1 % | 152 | 2 % | 36 % | 341 | 2 % | 47 % |
| camera_surveillance | 0 | 0 % | 288 | 3 % | 67 % | 1,497 | 9 % | 94 % |
| drone_monitoring | 0 | 0 % | 7 | 0 % | 19 % | 159 | 1 % | 59 % |
| bus_monitoring | 0 | 0 % | 1 | 0 % | 0 % | 26 | 0 % | 34 % |
| **Total** | **678** | **100 %** | **8,895** | **100 %** | **26 %** | **16,207** | **100 %** | **33 %** |

*Table B.3 Daily median device (and application) density*

|  | 2019 | | 2030 - slow demand scenario | | | 2030 - rapid demand scenario | | |
|---|---|---|---|---|---|---|---|---|
| Traffic category and device | density [device/km2] | share | density [device/km2] | share | CAGR | density [device/km2] | share | CAGR |
| **High-priority traffic** | **0** | **0 %** | **3** | **0** | **11 %** | **24** | **0 %** | **33 %** |
| bus_remote_driving | 0 | 0 % | 3 | 0 | 11 % | 24 | 0 % | 33 % |
| **Human traffic** | **20,413** | **93 %** | **22,466** | **1** | **1 %** | **22,503** | **62 %** | **1 %** |
| smartphone_all | 18,016 | 82 % | 19,775 | 1 | 1 % | 19,775 | 55 % | 1 % |
| modem_all | 2,344 | 11 % | 2,388 | 0 | 0 % | 2,388 | 7 % | 0 % |
| car_infotainment | 47 | 0 % | 268 | 0 | 17 % | 301 | 1 % | 18 % |
| **Machine low-activity traffic** | **1,409** | **6 %** | **4,366** | **0** | **11 %** | **12,890** | **36 %** | **22 %** |
| urban_sensors* | 0 | 0 % | 391 | 0 | 72 % | 1,629 | 5 % | 96 % |
| wearables | 62 | 0 % | 2,323 | 0 | 39 % | 6,337 | 18 % | 51 % |
| connected_bikes | 0 | 0 % | 31 | 0 | 37 % | 66 | 0 % | 46 % |
| pos_devices | 1,078 | 5 % | 1,078 | 0 | 0 % | 4,311 | 12 % | 13 % |
| smart_meters | 269 | 1 % | 539 | 0 | 7 % | 539 | 1 % | 7 % |
| **Machine high-activity traffic** | **47** | **0 %** | **392** | **0** | **21 %** | **689** | **2 %** | **28 %** |
| camera_surveillance | 0 | 0 % | 102 | 0 | 52 % | 205 | 1 % | 62 % |
| car_monitoring | 47 | 0 % | 268 | 0 | **17 %** | 301 | **1 %** | **18 %** |
| drone_monitoring | 0 | 0 % | 12 | 0 | 25 % | 119 | 0 % | 54 % |
| bus_monitoring | 0 | 0 % | 3 | 0 | 11 % | 24 | 0 % | 33 % |
| **Total** | **21,876** | **100 %** | **27,257** | **1** | **2 %** | **36,112** | **100 %** | **5 %** |

*urban sensors also include traffic lights and signs

*Table B.4 Peak hour device (and application) density*

| Traffic category and device | 2019 | | 2030 - slow demand scenario | | | 2030 - rapid demand scenario | | |
|---|---|---|---|---|---|---|---|---|
| | density [device/km2] | share | density [device/km2] | share | CAGR | density [device/km2] | share | CAGR |
| **High-priority traffic** | **0** | **0 %** | **3** | **0 %** | **11 %** | **24** | **0 %** | **34 %** |
| bus_remote_driving | 0 | 0 % | 3 | 0 % | 11 % | 24 | 0 % | 34 % |
| **Human traffic** | **37,699** | **96 %** | **41,415** | **86 %** | **1 %** | **41,448** | **68 %** | **1 %** |
| smartphone_all | 35,307 | 90 % | 38,755 | 80 % | 1 % | 38,755 | 63 % | 1 % |
| modem_all | 2,344 | 6 % | 2,388 | 5 % | 0 % | 2,388 | 4 % | 0 % |
| car_infotainment | 48 | 0 % | 272 | 1 % | 17 % | 306 | 0 % | 18 % |
| **Machine low-activity traffic** | **1,469** | **4 %** | **6,592** | **14 %** | **15 %** | **18,965** | **31 %** | **26 %** |
| urban_sensors* | 0 | 0 % | 391 | 1 % | 72 % | 1,629 | 3 % | 96 % |
| wearables | 122 | 0 % | 4,552 | 9 % | 39 % | 12,420 | 20 % | 51 % |
| connected_bikes | 0 | 0 % | 32 | 0 % | 37 % | 67 | 0 % | 47 % |
| pos_devices | 1,078 | 3 % | 1,078 | 2 % | 0 % | 4,311 | 7 % | 13 % |
| smart_meters | 269 | 1 % | 539 | 1 % | 7 % | 539 | 1 % | 7 % |
| **Machine high-activity traffic** | **48** | **0 %** | **401** | **1 %** | **21 %** | **768** | **1 %** | **29 %** |
| camera_surveillance | 0 | 0 % | 102 | 0 % | 52 % | 205 | 0 % | 62 % |
| car_monitoring | 48 | 0 % | 272 | **1 %** | **17 %** | 306 | **0 %** | **18 %** |
| drone_monitoring | 0 | 0 % | 23 | 0 % | 33 % | 233 | 0 % | 64 % |
| bus_monitoring | 0 | 0 % | 3 | 0 % | 11 % | 24 | 0 % | 34 % |
| **Total** | **39,216** | **100 %** | **48,411** | 100 % | 2 % | **61,205** | 100 % | 4 % |

*urban sensors also include traffic lights and signs